\documentclass[sigconf]{acmart}
\usepackage{xcolor}
\usepackage{listings}
\usepackage{makecell}
\newcommand{\ignore}[1]{}
\usepackage{ifpdf}

\usepackage[title]{appendix}

\usepackage{algorithmic}
\usepackage{array}
\usepackage{mdwmath}
\usepackage{mdwtab}
\usepackage{eqparbox}
\usepackage[tight,footnotesize]{subfigure}
\usepackage[]{caption}
\usepackage{fixltx2e}
\usepackage{stfloats}
\usepackage{url}

\usepackage{booktabs} 
\usepackage{tabu}
\usepackage{graphicx}
\usepackage{subfigure}
\usepackage{multirow}
\usepackage{pifont}
\newcommand{\xmark}{\ding{55}}%
\def\BibTeX{{\rm B\kern-.05em{\sc i\kern-.025em b}\kern-.08emT\kern-.1667em\lower.7ex\hbox{E}\kern-.125emX}}

\begin{document}
	
	%
	\title{Your Smart Home Can't Keep a Secret: \\
		Towards Automated Fingerprinting of IoT Traffic with Neural Networks}

	
	\newcommand\cjy[1]{\{\textbf{JC:} \textcolor{red}{\em#1}\}}
	\newcommand\zl[1]{\{\textbf{ZL:} \textcolor{blue}{\em#1}\}}
	\newcommand\dsk[1]{\{\textbf{dsk:} \textcolor{purple}{\em#1}\}}
	\newcommand\td[1]{\{\textbf{TD:} \textcolor{orange}{\em#1}\}}
	\newcommand\zkh[1]{\{\textbf{ZKH:} \textcolor{green}{\em#1}\}}
	\newcommand{\system}{\texttt{HomeMole}}
	%
    \author{Shuaike Dong}
    \email{ds016@ie.cuhk.edu.hk}
    \affiliation{The Chinese University of Hong Kong}
    \author{Zhou Li}
    \email{zhou.li@uci.edu}
    \affiliation{University of California, Irvine}
    
    \author{Di Tang}
    \email{td016@ie.cuhk.edu.hk}
        \affiliation{The Chinese University of Hong Kong}

    \author{Jiongyi Chen}
    \email{cj015@ie.cuhk.edu.hk}
        \affiliation{The Chinese University of Hong Kong}

    \author{Menghan Sun}
    \email{sm017@ie.cuhk.edu.hk}
        \affiliation{The Chinese University of Hong Kong}

    \author{Kehuan Zhang}
    \email{khzhang@ie.cuhk.edu.hk}
    \affiliation{The Chinese University of Hong Kong}
    %
    %
    
	\begin{abstract}
	The IoT (Internet of Things) technology has been widely adopted in recent years and has profoundly changed the people's daily lives. However, in the meantime, such a fast-growing technology has also introduced new privacy issues, which need to be better understood and measured. 

In this work, we look into how private information can be leaked from network traffic generated in the smart home network. 
Although researchers have proposed techniques to infer IoT device types or user behaviors under clean experiment setup, the effectiveness of such approaches become questionable \textit{in the complex but realistic network environment}, where common techniques like Network Address and Port Translation (NAPT) and Virtual Private Network (VPN) are enabled. Traffic analysis using traditional methods (e.g., through classical machine-learning models) is much less effective under those settings, as the features picked manually are not distinctive any more.

In this work, we propose a traffic analysis framework based on sequence-learning techniques like LSTM and leveraged the temporal relations between packets for the attack of device identification. 
We evaluated it under different environment settings (e.g., pure-IoT and noisy environment with multiple non-IoT devices). The results showed our framework was able to differentiate device types with a high accuracy. This result suggests IoT network communications pose prominent challenges to users' privacy, even when they are protected by encryption and morphed by the network gateway. As such, new privacy protection methods on IoT traffic need to be developed towards mitigating this new issue.

	\end{abstract}
	
	%
	%

	
	%
	
	%
	
	%
	\maketitle
	
	\section{Introduction}
\label{sec:introduction}

The Internet of Things (IoT) have been gaining popularity increasingly in recent years, and continue to expand in areas such as smart homes, smart cities, industrial systems, connected health products, and so on. According to reports from Forbes, the global IoT market will grow from 157 billion US dollars in 2016 to 457 billion US dollars by 2020, attaining a Compound Annual Growth Rate (CAGR) of 28.5\%~\cite{url_forbes}. 


Smart home is a prominent use case of IoT, where multiple IoT devices work together to facilitate all kinds of user activities by sensing surroundings, interpreting human commands and providing feedback. However, smart home can introduce new threats to residents' privacy. Since network packets between IoT device and remote server can be eavesdropped, a motivated attacker can leverage such data to infer private information about the residents, like what IoT devices are installed and whether they are active. Leaking such information would cause grave consequences to the residents: e.g., a theft can break into the home when no one is inside by learning the status of the installed smart switch.


This paper aims to assess the privacy threat to smart home residents by evaluating different traffic-analysis approaches on datasets carrying real IoT traffic.
Though a few recent works also investigated the privacy issues related to IoT network communications~\cite{DBLP:journals/corr/ApthorpeRF17,DBLP:conf/infocom/SivanathanSGRWV17,Acar2018PeekaBooIS}, those works all assume a local adversary (only the traffic between IoT device and gateway can be sniffed) or relatively simple network environment (e.g., traffic from devices can be easily separated). Whether traffic analysis is effective under a remote adversary or more complex network is not yet assessed. 
In particular, we assume that the the gateway may enable configurations that are common but hamper traffic analysis, like Virtual Private Network (VPN) and Network Address and Port Translation (NAPT). Under those settings, traffic flows belonging to different IoT devices could be merged to a single flow and the valuable information from fields like destination ports could be erased. Still, our study shows by exploiting the temporal relations between packets of an individual device, the device can be reliably identified.

More specifically, we found such temporal relations can be modeled by sequence model LSTM-RNN when grouping consecutive packets into a traffic window. We carefully designed the structure of both models and evaluated on two datasets filled with traffic generated from off-the-shelf IoT devices and non-IoT devices. The evaluation results show our models can achieve better accuracy comparing to the models widely used by existing works, like Random Forest. To highlight a few, our bidirectional LSTM model can achieve an accuracy of 99.2\% and 97.7\% on IoT devices in NAPT and VPN configuration. Even when a large-amount of non-IoT traffic are generated at the same time, it can still achieve 92.1\% and 81.0\% accuracy in these two configurations.

\noindent\textbf{Contributions.}
We summarize the technical contributions made by this work as below.
\begin{enumerate}
    
\item We present a traffic-analysis system, \system, to automatically infer the IoT devices behind a smart home network even when traffic fusion like NAPT and VPN are enabled. We designed a basic LSTM model and a  bidirectional LSTM model that are able to identify IoT devices based on the sniffed packets.

\item We evaluated our system under two network configurations (NAPT and VPN) and two scenarios (pure-IoT and noisy environment). The results indicate that our framework could achieve high accuracy under those settings. Our models outperform the baseline model, Random Forest, due to the ability of modeling the temporal relations between packets.
  

\end{enumerate}

To facilitate the research in this domain, we will release our datasets, together with models we built. We believe by releasing datasets and models, we could help other researchers to investigate new traffic-analysis methods and IoT community to build better defense.

\noindent\textbf{Paper organization.}
The paper is organized as follows: Section~\ref{sec:background} presents the background about the relevant previous work, smart home network, our adversary model and neural networks used in our paper. Section~\ref{sec:system} describes the insights into IoT traffic, and the design of our framework. Section~\ref{sec:evaluation} presents the experimental results of our framework in different scenarios. Section~\ref{sec:discussion} discusses the limitations of this work. In the end, we conclude this study in Section~\ref{sec:conclusion}. 

	\section{Background}
\label{sec:background}

\subsection{Related Work}
\vspace{2pt} \noindent
\textbf{Network traffic classification.} 
Network traffic analysis has been shown reasonably effective when applied on anomaly detection ~\cite{DBLP:journals/corr/HuangAB14,DBLP:conf/ccis/YuLZY14}, software identification~\cite{DBLP:conf/ndss/PanchenkoLPEZHW16,DBLP:conf/ndss/VeraRimmer2018,DBLP:conf/cec/AksoyLG17}, individual user fingerprinting~\cite{DBLP:conf/sigcomm/VassioGTMS17,DBLP:conf/icdcs/VerdeAGMS14} and etc. At a high level, the techniques used in existing applications fall into the following two directions: deep packet inspection and side-channel inference.

Deep packet inspection(DPI) is effective for packet classification and intrusion detection when the network traffic is unencrypted. Bujlow et al. \cite{DBLP:journals/cn/BujlowCB15} conducted a comparison among 6 well-known DPI tools and found that the commercial DPI tool achieves very good performance in traffic classification. However, DPI-based approaches become ineffective when it applied on encrypted traffic. A lot of works have been proposed to analyze such traffic~\cite{DBLP:conf/ndss/WrightCM09, DBLP:conf/sp/DyerCRS12, DBLP:conf/sp/SunSWRPQ02, DBLP:conf/pet/BissiasLJL05, DBLP:conf/ccs/LiberatoreL06} by utilizing the side-channel information and metadata, like source and destination IP address, port and packet size. Apart from that, previous works like~\cite{DBLP:conf/huc/SrinivasanSW08} also show other side-channel information, such as wireless signal strength and timing, can be leveraged to infer user activities. 
Machine learning based techniques have shown many successes for traffic classifications. For example, Taylor et al. \cite{DBLP:conf/eurosp/TaylorSCM16} proposed an identification framework for smartphone apps called \texttt{AppScanner}, which extracts statistical features from network flows for classification tasks. Trained by different learning algorithms, AppScanner can reach a highest accuracy of 99.8\% on 110 Apps. Chen et al.~\cite{DBLP:conf/bigdataconf/ChenHLG17} proposed an online traffic classification framework which utilizes kernel methods and deep neural networks. They tested their tool on 5 different protocols and 5 mobile applications and achieved 99.84\% and 88.43\% correspondingly. However, those two approaches only works on relatively simple network environment where none of NAT or VPN is enabled. 

\vspace{2pt} \noindent
\textbf{Traffic classification in IoT domain.} 
Following the rapid development of the IoT ecosystem, how to characterize and fingerprint of IoT devices has become a trending topic. 
There have been some works on IoT traffic analysis~\cite{DBLP:conf/infocom/SivanathanSGRWV17, DBLP:journals/corr/abs-1708-05044, DBLP:journals/corr/SibyMT17, DBLP:journals/access/MartinCSL17, DBLP:journals/corr/ApthorpeRF17, miettinen2017iot} proposed in recent years. 
Siby et al.~\cite{DBLP:journals/corr/SibyMT17} captured radio signals emitted from IoT devices and created a system to store and visualize the traffic. Recently,
Apthorpe et al.~\cite{DBLP:journals/corr/ApthorpeRF17} performed case studies on four IoT devices, and show they exhibit distinctive traffic features, which could enable device identification and behavior inference.  Marcus et al. ~\cite{miettinen2017iot} extracted 23 features from raw packets including network protocols from Link layer to Application layer,  IP options, IP addresses and ports. However, part of their features are inaccessible in our settings due to the adversary capabilities and complex network configurations (NAPT, VPN). For example, ARP and LLC are only capturable in local area network environment, SSDP and MDNS protocols account for only small amount of traffic that cannot holistically describe the network status. Compared with them, our model only takes advantage of 10 features extracted from packet metadata which are illustrated elaborately in Section~\ref{sec:system}. Sivanathan et al.~\cite{DBLP:conf/infocom/SivanathanSGRWV17} collected network traces of more than 20 IoT devices in a campus environment over 3 weeks and characterized the profiles of those IoT devices according to their traffic patterns. It also relies on an extensive feature engineering to select the salient features and some of them become vain when a complex network setting is applied, like NAPT and VPN.


\begin{figure}[t]
	\centering
	\includegraphics[width=0.85\columnwidth]{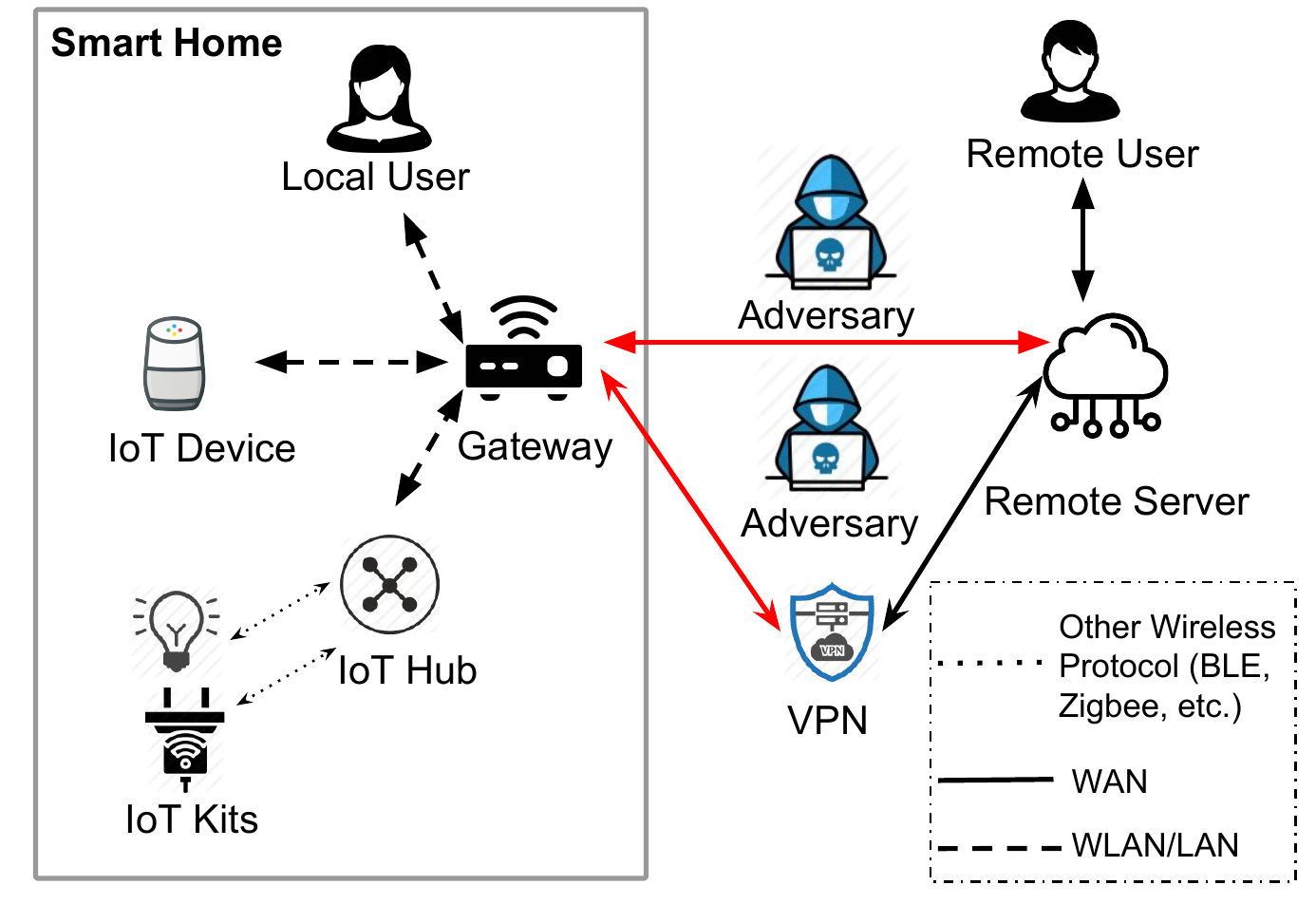}     
	\caption{Network structure of a typical smart home.}
	\label{fig:home_net}
	\vspace{-0.2in}
\end{figure}

\subsection{Smart Home Network}
\label{sec:smartnetwork}
We assume the network communication within a smart home involves four parties: IoT device, service provider, gateway and the user. The communication schema is illustrated in Figure~\ref{fig:home_net}. Below we briefly overview their communication schema.
\begin{itemize}
\item The first type of \textit{IoT device} senses the surrounding environment and sends notifications to its associated listeners, either periodically or immediately when the event takes place. For example, Samsung ST Motion Sensor detects when a person approaches in proximity~\cite{url_samsungmotionsensor} and notifies other IoT devices. Another type of IoT device is \textit{IoT hub}, which acts as the ``brain'' (centroid controller) for other IoT devices (or \textit{IoT kits}) in the close range. IoT hub is necessary in order to control IoT devices that use low-power protocols like BLE. Both regular IoT device and IoT hub can be controlled by user's commands sent remotely or locally.

\item The IoT device interacts with the \textit{service provider} operated by its manufacturer through Internet communications. The service provider is responsible for request handling and resource relaying.
To reduce the operational cost, many device vendors have moved their services to the public cloud infrastructure and leverage the cloud analytics, like AWS IoT Core~\cite{url_awsiotcore}, to process massive amount of IoT data.

\item A \textit{gateway} is a bridge between in-home IoT devices and the remote service provider. A typical gateway (e.g., router) supplies two types of interfaces for in-home devices, WLAN and LAN. The communication with service provider is through WAN interface.

\item A \textit{user} takes control of all smart devices in his/her home. There are usually two ways to interact with IoT devices, through mobile applications or human interactions (e.g., walking, talking, and touching). 

\end{itemize}



\noindent\textbf{Device identifier.} 
IoT devices within the smart home network can be distinguished by device identifiers determined by specific network protocols. Each Wi-Fi device has a unique source IP address and MAC address. Though IoT kits might not get IP addresses in smart home if they are not using WiFi, they can still obtain identifiers from IoT hub through other protocols (e.g., NwkAddr for Zigbee devices and Resolvable Private Address for BLE devices). 

\subsection{NAPT and VPN}
\label{sec:tunneling}

When the network traffic is observed between the gateway and the service provider, the original device identifiers may be obscured. Firstly, the gateway could strip off the source MAC address from the packets~\cite{macremoval}. Even the port information can be changed under NAPT (Network Address and Port Translation) or VPN (Virtual Private Network). Below, we describe how changes are made by NAPT and VPN.


\noindent \textbf{NAPT}. 
To conserve the limited global IPv4 space, NAPT is developed to enable the sharing of one IP address among different devices.
In particular, NAPT modifies the network-layer and transport-layer identifiers like destination IP address and destination port numbers of inbound packets\cite{DBLP:journals/rfc/rfc2663}. For outbound packets, the source IP address and the source port are translated. In both cases, the IP address of a local device is replaced with the gateway's IP address. 
The gateway using NAPT holds a \textit{translation table} which records the mapping of addresses and ports so that packets will be routed to the right destination.


    

    
\begin{figure}[t]
	\centering
	\includegraphics[width=0.85\columnwidth]{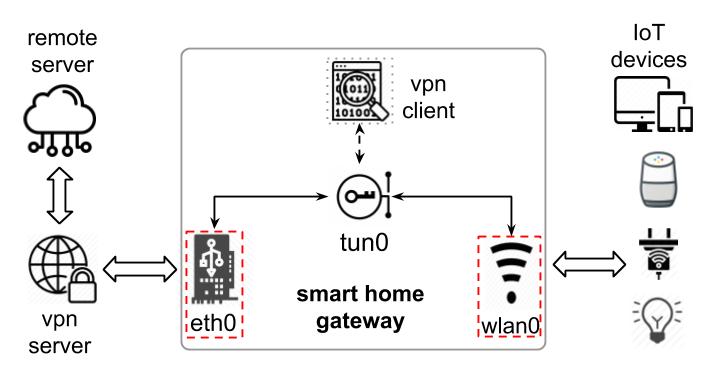}           
	\caption{VPN-enabled gateway.}
	\label{fig:vpn_structure}
	\vspace{-0.2in}
\end{figure}

\noindent\textbf{VPN}. 
VPNs are often used to interconnect different networks to form a new network with a larger capacity~\cite{DBLP:journals/rfc/rfc2764}. Based on the IP tunneling mechanism, hosts in different subnets can communicate with each other and the delivered information can be kept secret with authentication and encryption. 
Figure~\ref{fig:vpn_structure} shows the network structure after deploying a VPN-enabled gateway. Different from normal routers, a VPN-enabled gateway owns three network interfaces -- \texttt{wlan0}, \texttt{eth0} and \texttt{tun0}. Among them, \texttt{wlan0} works as the entrance of LAN, collecting and delivering packets from local devices.
The Ethernet interface \texttt{eth0} holds the connection between the gateway and WAN.
\texttt{tun0} is created by VPN client process. 
For every packet from \texttt{wlan0} to \texttt{eth0}, the VPN client first encrypts the original packet into a payload and constructs a new packet. The new packet is then delivered to a VPN server and gets decrypted. The VPN server then forwards restored packets to their original destinations. 
From the viewpoint of destination remote server, the original metadata like source IP and source port are completely hidden, which protects user's privacy against on-path eavesdroppers.

    
\subsection{Adversary Model}
\label{sec:adversary}

The goal of our adversary is to identify the \textit{active} IoT devices in the targeted smart home. Such inference can leak sensitive information in a smart home. For example, an ISP can infer the device information and sell them to advertisers who like to do targeted advertising. Or a theft can pick the time when the user is not at home by sniffing the outbound traffic and inferring the status of the installed surveillance camera. To this end, in this paper we consider passive eavesdroppers who can observe the \textit{encrypted} network traffic flowing \textit{between the gateway and the remote service}. 
More importantly, there are two realistic settings that are not considered by prior works.
On one hand, we assume that NAPTs or VPNs are enabled in the gateway so that the original device identifiers are replaced by the gateway's and the traffic belonging to different devices are \textit{merged}, even the contacted remote server becomes opaque to the adversary.
On the other hand, we assume multiple devices (including IoT devices and non-IoT devices such as mobile phones and tablets) may work \textit{simultaneously} such that their packets are interleaving. As shown in ~\cite{DBLP:conf/infocom/SivanathanSGRWV17}, non-IoT devices usually have a higher rate in generating packets and their volume is larger than that of IoT devices, which means the existence of non-IoT devices can significantly distort the original statistical features learned on IoT devices.

Previous works~\cite{DBLP:journals/corr/abs-1708-05044, Acar2018PeekaBooIS, DBLP:conf/eurosp/TaylorSCM16} assumed that the adversary can sniff traffic within the smart home network (i.e., \textit{local adversary}) or traffic fusion is not performed by the gateway. In their scenarios, the flows from different devices are clearly separated based on device identifiers. Unlike previous works, the remote adversary in our study is more realistic and the traffic analysis is much more challenging to perform.


\ignore{
The adversary is very common in daily life. \dsk{some news and reports?} For example, some network operators like ISP and the government can act as the role, so are the network department in a company, who is able to monitor the network activities occurring at the sub-gateways.
}
\ignore{
As a result, information like the domains visited by IoT devices might not be available: the gateway could cache DNS resolution result before the adversary starts to monitor the communication and flush out the cache after a long period, so the domain names of the service providers would be absent for traffic analysis, which have been leveraged by previous works~\cite{DBLP:journals/corr/abs-1708-05044}.
}
\subsection{LSTM-RNN}
\label{sec:lstm_rnn}

\ignore{
Previous works on traffic analysis mainly focus on selecting the most salient features  through extensive feature engineering~\cite{DBLP:conf/infocom/SivanathanSGRWV17,DBLP:journals/corr/ApthorpeRF17}. Such hand-picked features may require long collection time (e.g. feature like average sleep time) or be easily distorted when different devices are active in the same period of time (e.g. feature like mean traffic rate). 
}

Recently, deep neural network (DNN) has been gaining traction in the security domain and shown many promising results, given its capability of feature representation learning. For example, the research of Rimmer et al. demonstrated that websites visited by Tor users can be fingerprinted automatically with DNN~\cite{DBLP:conf/ndss/VeraRimmer2018}. 

\begin{figure}[t]
	\centering
	\includegraphics[width=0.85\columnwidth]{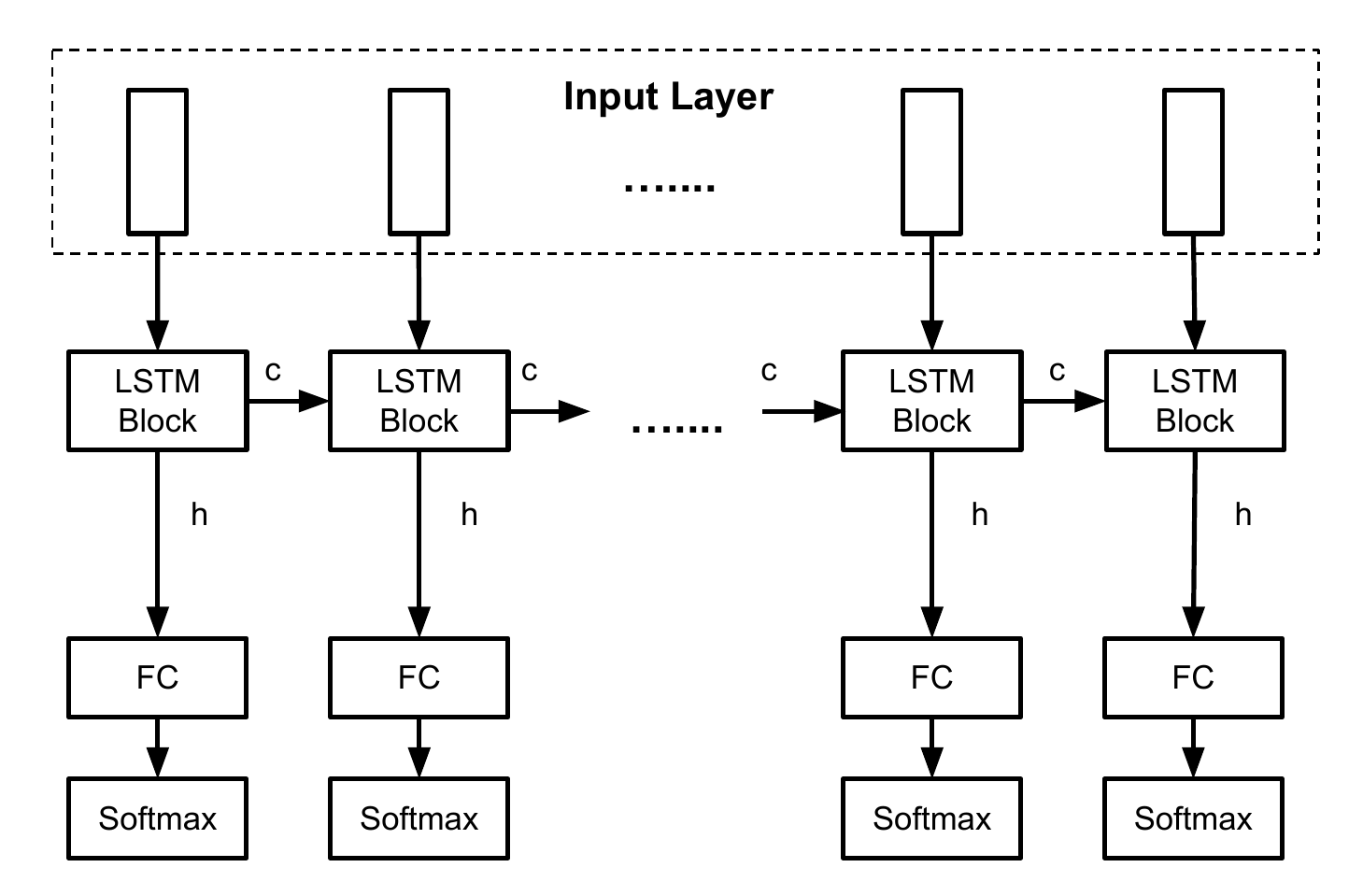}
	\caption{Structure of a basic LSTM used in our work (FC means fully-connected layer).}
	\label{fig:lstm}
	\vspace{-0.2in}
\end{figure}

Recurrent Neural Network (RNN) is one type of DNN that is good at handling temporal-related sequences. With multiple recurrent cells connected, the output of a previous cell can be passed to the current one. In this way, historical information is kept and forwarded.
Among different implementations of RNN, LSTM(Long Short-Term Memory)-RNN has become a popular choice as it is able to address weakness of other RNNs like exploding and vanishing gradient. 
It provides a novel memory cell consisting of three different gates: input gate, forget gate and output gate. These gates are used to process the data transferred from previous memory cell and manipulate the current cell state. 
LSTM-RNN has achieved many successes in different areas, such as speech recognition~\cite{Graves2013HybridSR}, medical diagnose~\cite{Lipton2016LearningTD} and system log analysis~\cite{DBLP:conf/ccs/Du0ZS17}. Figure~\ref{fig:lstm} shows the structure of a basic LSTM-RNN.


Inspired by the recent research, we find our problem is a natural fit for LSTM-RNN models. Similar to system logs, traffic generated by IoT devices can be organized in chronological order. There exists contextual dependency between packets based on the running states of the device and such dependency can be modeled by LSTM-RMM models.
In Section~\ref{sec:system}, we describe our LSTM-RNN models in details.

	\section{System Design}
\label{sec:system}

In this section, we elaborate the design and implementation of our system, named \system, for fingerprinting IoT traffic. The goal is to identify the active IoT devices in a smart-home environment based on their network traffic. Different from those doing fingerprinting at flow level\cite{DBLP:conf/eurosp/TaylorSCM16,DBLP:conf/bigdataconf/ChenHLG17}, \system\  works at \textit{packet} level, which means all the packets will be given a label after processed by our models. 
As such, \system\ is able to work in \textit{online} mode and give prompt results. In addition to device identification, \system\ can also be used for other scenarios like QoS (Quantity of Service) and IDS (Intrusion Detection System), as shown in~\cite{fraleigh2003packet, DBLP:conf/ndss/MirskyDES18}. 

Below, we first describe our insights into IoT traffic analysis in Section~\ref{sec:observation}. Next, we elaborate how we set up the environment to collect data in Section~\ref{sec:experiment_setup}. After that, we explain how we prepare datasets from raw collected packets in Section~\ref{sec:traffic_preprocess}. Finally, we show the detailed structure of our models and why they are adequate for smart-home environment in Section~\ref{sec:model_select}. 

\subsection{Insights into IoT traffic} 
\label{sec:observation}


We carried out exploratory analysis on the realworld IoT devices and public dataset to characterize their network communication patterns. We identify several insights which highlight the uniqueness of IoT traffic comparing to the desktop and mobile traffic. 

\noindent 1) \textit{The devices belong to the same category have similar traffic patterns}.
As one example, we show the traffic patterns of Amazon Echo Dot and Google Voice Assistant (both are voice assistants) when they are waken up (see Figure~\ref{fig:echodot} and Figure~\ref{fig:google_voice} in Appendix~\ref{sec:traffic_pattern}).
As can be seen, when the voice command is recognized, both of them generate a traffic burst, followed by a period of continuous communication with the remote server. 
        

\noindent 2) \textit{Heartbeat communication is prevalent}. The service provider relies on the heartbeat messages sent by the IoT device to acquire liveness of the device. The heartbeat packets usually have constant size and interval, and different devices tend to use different modes for heartbeat. For example, Amazon Echo Dot sends a 95-byte heartbeat packet in TLSv1.2 format every 30 seconds, and Google Home sends two consecutive heartbeat packets in TLSv1.2 format every 60 seconds with the size of 135 bytes and 104 bytes. In addition, heartbeat communications are rarely disturbed by other activities happening on the devices. As such, they are good indicators to distinguish the categories or even manufacturers. 
    
\noindent 3) \textit{Protocol choices are diverse.} The communication between devices and remote services must follow certain convention, which is reflected in the chosen protocols. 
In Table~\ref{tab:protoc_statis} we show such diversified choices in terms of packets ratio under different network protocols.
Note that we only measure protocols at network layer, transmission layer and application layer, discarding those from lower layers.
According to our statistics, most traffic captured by an attacker is \texttt{IPv4}-based. For devices with large volumes of traffic, like network camera, \texttt{UDP} is usually adopted. Compared with non-IoT devices, IoT devices generate much less \texttt{HTTP} traffic for communication. Though \texttt{DNS} is usually used by previous works for device identification~\cite{DBLP:journals/corr/ApthorpeRF17, DBLP:conf/infocom/SivanathanSGRWV17, DBLP:journals/corr/abs-1708-05044}, our results show that its ratio is quite low compared to other protocols. As such, simply relying on \texttt{DNS} does not guarantee device identification especially when the traffic observed comes from an incomplete session.
\begin{table}[]
	\scalebox{0.85}[0.9]{%
	\begin{tabular}{|c|c|c|c|c|c|c|c|}
		\hline
		\textbf{Device (\%)}                                            & \textbf{IPv4} & \textbf{UDP} & \textbf{TCP} & \textbf{TLS} & \textbf{HTTP} & \textbf{DNS} & \textbf{O}                                     \\ \hline
		Google Home                                                   & 100.0         & 1.5          & 98.1         & 26.6         & 0             & 0.6          & 0.4                                                          \\ \hline
		Echo Dot                                                      & 100.0         & $\sim$0      & $\sim$100.0  & 14.6         & $\sim$0       & $\sim$0      & $\sim$0                                                     \\ \hline
		Tmall Assist                                                & 99.6          & 0            & 99.6         & 21.0         & 5.5           & $\sim$0      & 0.4                                                           \\ \hline
		\begin{tabular}[c]{@{}l@{}}360 Cam\\ (LAN mode)\end{tabular} & 99.7          & 78.6         & 21.1         & 0.4          & 0             & 0            & 0.4                                                          \\ \hline
		\begin{tabular}[c]{@{}l@{}}360 Cam\\ (WAN mode)\end{tabular} & 100.0         & 99.9         & 0.1          & $\sim$0      & 0             & 0            & 0                                                            \\ \hline
		Orvibo                                                     & 99.6          & 0.2          & 99.4         & $\sim$0      & 0             & 0            & 0.4                                                         \\ \hline
		Broadlink                                                  & 99.7          & 99.7         & 0            & 0            & 0             & 0            & 0.3                                                             \\ \hline
		Tplink                                                    & 99.4          & 0.1          & 99.3         & 50.9         & 0             & 0            & 0.6                                                            \\ \hline
		Xiaomi Hub                                                      & 99.5          & 99.5         & 0            & 0            & 0             & 0            & 0.5                                                               \\ \hline
		Noise - mobile                                                  & 87.6          & 5.5          & 80.1         & 4.5          & 1.7           & 2.2          & 12.4             \\ \hline
		Noise - tablet                                                  & 87.0          & 0.5          & 86.4         & 0.6          & 0.1           & 0.1          & 13.0                                                       \\ \hline
	\end{tabular}
	}
    \caption{Protocol distribution (\textbf{O} means other protocols).}
\label{tab:protoc_statis}
\end{table}


In short, the above observations suggest fingerprinting IoT devices is feasible, even under complex network environment like NAPT and VPN. On the other hand, a comprehensive model instead of matching individual feature is necessary for our task.

\ignore{
\begin{table}[t]
	\centering
	\caption{The proportion of shared-port traffic of IoT devices}
	\label{tab:proportion_unique_port}
	\begin{tabular}{|l|l|l|}
		\hline
		\textbf{Device} &
		\textbf{Proportion(total)} & \makecell{\textbf{Proportion}\\ \textbf{(no-port-packet-excluded)} }                                   
		
		\\ \hline
		iHome & 49.0\% & 99.8\%
		\\ 
		\hline
		Samsung SmartCam & 82.9\% &  99.5\%
		\\ 
		\hline
		Dropcam & 97.4\% & 100.0\%
		\\
		\hline
		Belkin wemo motion sensor & 43.7\% & 97.9\%
		\\
		\hline
		Withings Smart scale & 87.6\% & 100.0\%
		\\
		\hline
		Triby Speaker & 7.4\% & 19.7\%
		\\
		\hline
		Insteon Camera & 84.0\% & 98.4\%
		\\
		\hline
		NEST Protect smoke alarm &8.6\% & 10.2\%
		\\
		\hline
		Nest Dropcam & 99.0\% & 100.0\%
		\\
		\hline
		Light Bulbs LiFX Smart Bulb & 15.5\% & 21.1\%
		\\
		\hline
		Amazon Echo & 35.3\% & 56.3\%
		\\
		\hline
		Withings Smart Baby Monitor & 17.7\%  & 99.8\%
		\\
		\hline
		Netatmo Welcome & 46.7\%  &99.7\%
		\\
		\hline
		PIX-STAR Photo-frame & 46.1\% & 81.9\%
		\\
		\hline
		Withings Aura smart sleep sensor& 75.0\% & 99.9\%
		\\
		\hline
		Belkin Wemo switch & 52.1\% & 99.8\%
		\\
		\hline
		Netatmo weather station & 40.7\% & 80.7\%
		\\
		\hline
		Blipcare Blood Pressure meter & 18.3\% &24.6\%
		\\
		\hline
		
	\end{tabular}
\end{table}
}




\subsection{Data Collection}
\label{sec:experiment_setup}
While some prior works on IoT network analsyis have published their datasets~\cite{DBLP:conf/infocom/SivanathanSGRWV17}. We found they cannot be used in our study, as we focus on more complex settings with NAPT and VPN enabled in the gateway. As such, we set up our own smart home environment and collected the traffic by ourselves. The dataset will be published at a public repository.

We set up the environment in a campus laboratory with 15 devices, including 10 IoT and 4 non-IoT devices. Table ~\ref{tab:dev_list} shows the details of our devices. Our devices can be divided into six categories: voice assistant, IoT hub, IoT kits(smart plug), network camera, interactive machine and non-IoT deivces. Devices used in our paper have different interaction modes and traffic patterns, which can be helpful to depict the overall picture of IoT device traffic.

We use a Raspberry Pi~\cite{url_raspberry} as the gateway. A typical Raspberry Pi provides two network interface cards -- \texttt{eth0} and \texttt{wlan0}. To simulate NAPT, we connect \texttt{eth0} to the Internet and then enable the linux service \texttt{hostapd} to create an access point with \texttt{wlan0} network card of Raspberry Pi. Next, we create rules for \texttt{iptables} so that packets can be forwarded from \texttt{wlan0} to \texttt{eth0} and vice versa. For VPN, we establish a virtual machine with DigitalOcean Droplets service~\cite{url_digital_ocean} and use it as our VPN server, we then setup \texttt{openvpn} client on our Raspberry Pi to enable VPN tunneling.

\begin{table}[H]
\begin{tabular}{|c|c|c|}
\hline
\textbf{device}                                              & \textbf{MAC}      & \textbf{type}   \\ \hline
echo dot                                                     & 88:71:e5:ed:be:c7 & voice assistant \\ \hline
\begin{tabular}[c]{@{}l@{}}google home\end{tabular}        & f4:f5:d8:db:61:84 & voice assistant \\ \hline
\begin{tabular}[c]{@{}l@{}}tmall assist\end{tabular}    & 18:bc:5a:19:eb:7d & voice assistant \\ \hline
\begin{tabular}[c]{@{}l@{}}xiaomi hub\end{tabular}         & 78:11:dc:e1:f0:6b & hub             \\ \hline
\begin{tabular}[c]{@{}l@{}}360 camera\end{tabular}         & b0:59:47:34:16:ff & network camera  \\ \hline
\begin{tabular}[c]{@{}l@{}}xiaobai camera\end{tabular}     & 78:11:dc:cf:c8:f1 & network camera  \\ \hline
\begin{tabular}[c]{@{}l@{}}tplink plug\end{tabular}        & 30:20:10:fb:7c:05 & smart plug      \\ \hline
\begin{tabular}[c]{@{}l@{}}orvibo plug\end{tabular}        & b4:e6:2d:08:63:0c & smart plug      \\ \hline
\begin{tabular}[c]{@{}l@{}}broadlink plug\end{tabular}     & 78:0f:77:1b:00:8c & smart plug      \\ \hline
\begin{tabular}[c]{@{}l@{}}mitu story  teller\end{tabular} & 28:6c:07:87:54:b0 & interactive     \\ \hline
\begin{tabular}[c]{@{}l@{}}xiaomi mobile\end{tabular}      & a4:50:46:06:80:43 & non-IoT         \\ \hline
\begin{tabular}[c]{@{}l@{}}xiaomi tablet\end{tabular}      & 20:a6:0c:5a:42:10 & non-IoT         \\ \hline
\begin{tabular}[c]{@{}l@{}}sony mobile\end{tabular}        & 28:3f:69:05:2d:b0 & non-IoT         \\ \hline
\begin{tabular}[c]{@{}l@{}}motorola mobile\end{tabular}    & 44:80:eb:21:cb:95 & non-IoT         \\ \hline
\end{tabular}	
\caption{Devices used in our experiment}
	\label{tab:dev_list}
\end{table}


We collect network traffic under three different settings:
\begin{itemize}
	\item \textit{Single-device environment.} We assume only one device is active and we connect one IoT device to the gateway at a time.
	\item \textit{Multi-device and noisy environment.} In this case, all IoT and non-IoT devices are connected to the gateway which has NAPT enabled. Several devices may work simultaneously at time, leading to traffic fusion.
	\item \textit{VPN environment.} In addition to the above settings, we assume VPN is enabled. Traffic before and after VPN are both collected from \texttt{wlan0} and \texttt{eth0} at the same time.
\end{itemize}

To generate traffic, we adopt two strategies: automatic triggering and manual triggering. Automation can relieve the burden of tedious repeating experimenters and manual triggering can simulate human-machine interaction in real environments. 

\noindent \textbf{Automatic triggering.} For devices like smart plug and network that can be controlled by mobile apps, we use MonkeyRunner~\cite{url_monkeyrunner} to interact with the UI of mobile apps and trigger different functions of IoT devices. For devices like voice assistants directly controlled by human's input, we replay the commands near them. For example, Google Home plays songs when it hears the command ``sing a song''. We record a list of different commands and play them in a loop with a proper interval. As smart devices may have different responses even facing the same command, though the total amount of recorded commands is fixed, traffic we collect may vary.

\noindent \textbf{Manual triggering.} The Manual triggering is used in collecting traffic from multi-device scenario. In this setting, devices are set up in a shared room (laboratory),  people coming in and out this room can interact with the devices as they want. The functions triggered in this scenario and their time intervals are irregular and comply with what may happen in a real environment. Compared with the automatic triggering approach, manual triggering introduces more randomness to the dataset, which is helpful to the generalization of our models.

 \begin{figure}[t]
	\centering
	\includegraphics[width=1.0\columnwidth]{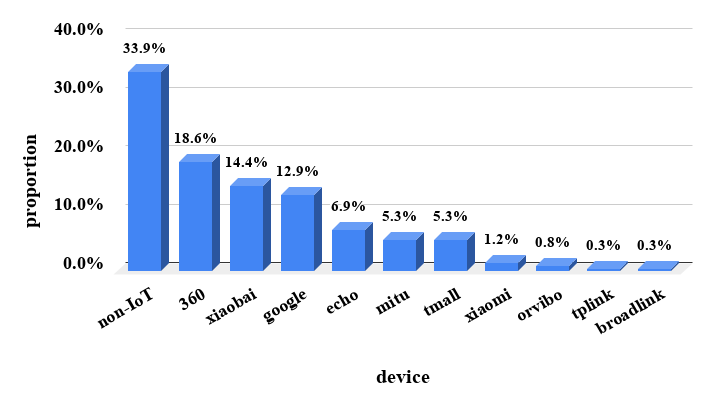} 
	\caption{Proportion of collected packets.}
	\label{fig:dataset_statis}
	\vspace{-0.2in}
\end{figure}

We adopt the popular network analysis tool \texttt{tshark} to monitor \texttt{wlan0} and \texttt{eth0} simultaneously. The traffic collected is dumped into files with extension ``.pcapng'' and is then pre-processed before classification.

The collection lasts for 49.4 hours. In a total, we collected 4.05 GB traffic with 7,223,282 available packets (those cannot be obtained by our adversary model are excluded, like packets only being transmitted inside the LAN). The distribution of packets is shown in Fig~\ref{fig:dataset_statis}. Note that, due to the internal functionalities of devices vary and users' different habits, the collected traffic does not comply with a uniform distribution. 

\ignore{
\begin{table*}[t]
	\centering
	\caption{Wireless Protocols}
	\label{tab:wireless_protocol_capture}
	\begin{tabular}{|l|l|l|}
		\hline
		\textbf{Type} & \textbf{Standards} & \textbf{Tool}                                                                             \\ \hline
		Wi-Fi & IEEE 802.11/a/b/g & AWUS036NH USB Adapter
		\\ 
		\hline
		BlueTooth & IEEE 802.15.1 & CC2540 USB Dongle
		\\ 
		\hline
		ZigBee & IEEE 802.15.4 & CC2531 USB Dongle
		\\
		\hline
	\end{tabular}
\end{table*}
}

\ignore{
\textbf{Collecting packets.} As described earlier in the adversary model, we define the Ethernet-based attack and the wireless-based attack. As such, we use a \texttt{Ethernet sniffer} and a \texttt{wireless sniffer}\zl{name of your tools}, to collect the Ethernet packets and the wireless packets respectively. For Ethernet packets, we are able to capture all the traffic between the entire smart home and the Internet using port mirroring on the gateway. 
Note that the packets that we captured do not contain source IP addresses because the home router (i.e. the gateway) enables the Network Address Translation (NAT) functionality to masquerade private IP address.
Apart from that, we also collect wireless traffic using wireless network interface controller (wireless NIC). The wireless NIC is switched to monitor mode~\cite{url_monitor_mode} to capture wireless packets in the air. 
Our traffic collector consists of two components, \texttt{gateway traffic collector} and \texttt{in-door traffic collector}. As figure ~\ref{fig:traffic_collector} shows, all the devices in our smart home environment are connected to LAN and WLAN interfaces of a Netcore v1.7 gateway. The WLAN interface of the router is connected to the public Internet. With NAPT function enabled, the gateway assigns different private ip addresses to the inner devices so that they can communicate with the remote servers. We then configure the port mirroring functionality ~\cite{url_port_mirroring} of the router so that all the traffic passing through its WAN and LAN interfaces will be mirrored to a certain interface. In our case, we mirror the traffic to LAN1 of the router, where a work station is connected. We run a tshark ~\cite{url_tshark} process on our work station that keeps collecting the packets flowing through the network adapter of the machine. 
}



\ignore{
\begin{itemize}
\item \textbf{Packets for establishing communication.} TCP handshake procedure~\cite{url_tcp_handshake} is designed to establish a reliable connection between two hosts. In the establishment procedure, the two parties will transmit 3 packets (SYN, SYN-ACK, ACK) sequentially. Typically all the handshake packets have the same sizes without providing any particular information about the connection they established. Therefore, we remove all the handshaking packets from our dataset and focus on packets that actually transmit contents (PSH packets).
\item \textbf{Packets for management and control.} IEEE 802.11 standard (Wi-Fi) defines three different frame types -- management frame, control frame and data frame. Management frames are used to establish and maintain communications. Control frames undertake the role of controlling the transmission of data frames. Similar to TCP handshake packets, these packets do not help us differentiate IoT devices. As such, we filter them out and focus on data frames which transmit device-specific and application-specific data.
\end{itemize}
}

\subsection{Traffic Pre-processing}
\label{sec:traffic_preprocess}
We utilize a multi-platform packet parsing framework called \texttt{PcapPlus Plus} to pre-process the traffic. The goal is to extract low-level but useful features from the packet and compose a numerical vector that can be processed by our models. 

\vspace{2pt} \noindent
\textbf{Feature selection.}
Due to the encryption enforced by the communication, we extract features from the metadata of packet headers. We select features from different layers -- \texttt{frame length} and \texttt{epoch time} from \texttt{physical layer}, and \texttt{destination port number} from \texttt{transport layer}. In addition, we use a binary sequence to represent the protocols in packet transmission. 
We select 6 most common protocol types including \texttt{IP}, \texttt{TCP}, \texttt{UDP}, \texttt{TLS/SSL}, \texttt{HTTP} and \texttt{DNS}, according to our measurement (see Table~\ref{tab:protoc_statis}). If a packet involves one of the protocols, the corresponding bit will be set to 1, otherwise 0. We set the last position of the `binary string' to be \texttt{others} for the protocols beyond the previous 6 protocols. 
For example, a UDP-based DNS request is represented as \texttt{<1010010>} and a NTP packet is represented as \texttt{<1010001>}. 
The only feature we consider beyond metadata is the \texttt{direction} of packet. We use 0 and 1 for inbound and outbound packets respectively. 

Note that we do not use the domain name in DNS response like previous works~\cite{DBLP:journals/corr/ApthorpeRF17, DBLP:journals/corr/abs-1708-05044} because DNS can be encrypted as well~\cite{DBLP:journals/rfc/rfc8484, Hu2016SpecificationFD}. 
The destination \texttt{IP} is not used because it is periodically changed when the IoT vendors run the remote server on public cloud, which has become a popular choice~\cite{url_awsiotcore}.

\begin{figure}[t]
	\centering
	\includegraphics[width=0.85\columnwidth]{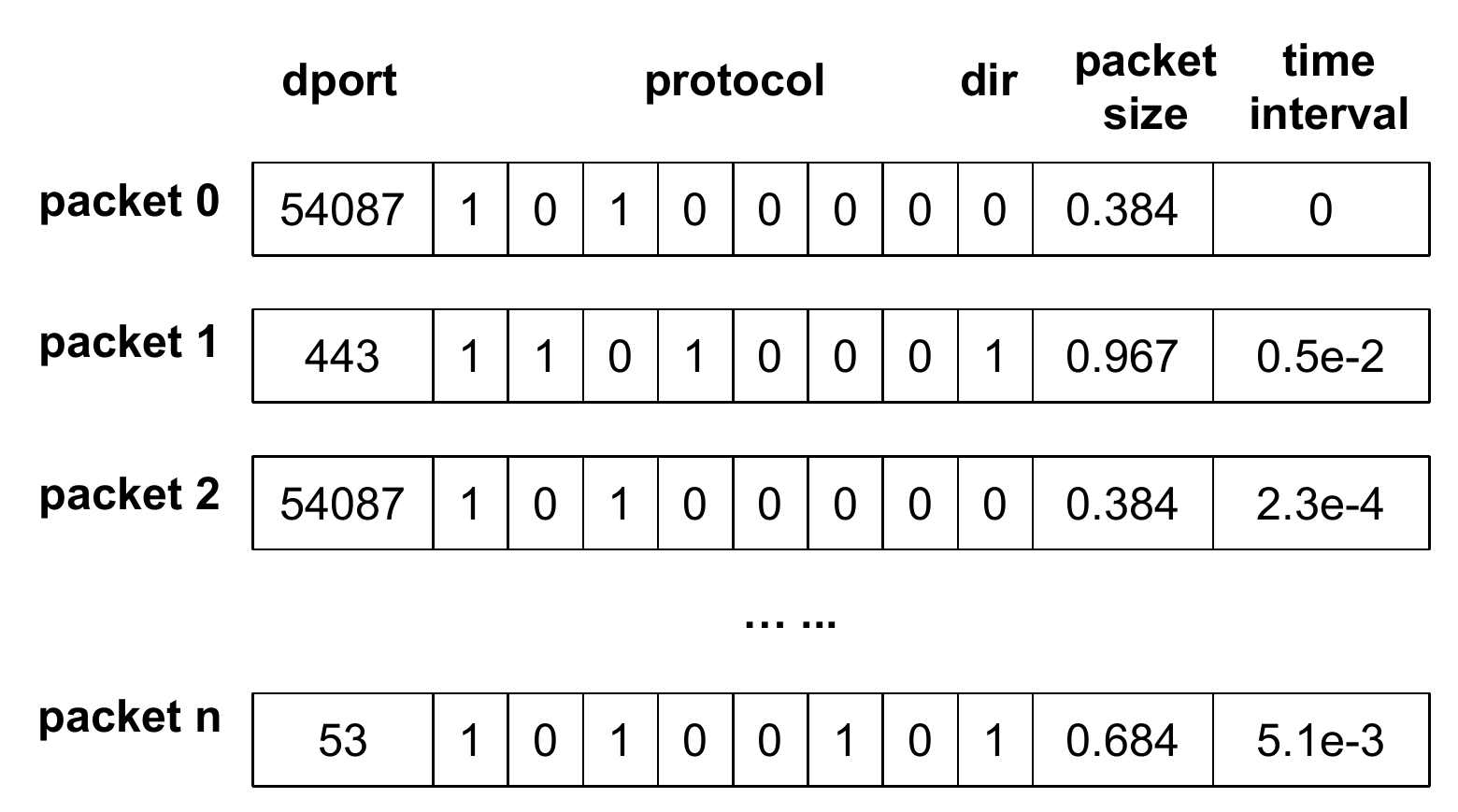} 
	\caption{An example of traffic window.}
	\label{fig:traffic_window}
	\vspace{-0.2in}
\end{figure}

In the end, we concatenate all the selected features and compose a one-dimensional vector as the representation of a packet (\texttt{<dport, protocol, direction, frame length, time interval>}), as shown in Figure~\ref{fig:traffic_window}. 
Note that we compute \texttt{time interval} from the  \texttt{epoch time} between two adjacent packets and use it as feature to model the temporal relations between packets.

\vspace{2pt} \noindent
\textbf{Packet labeling.} 
One key challenge in traffic pre-processing is packet labeling, especially under VPN environment. 
As section ~\ref{sec:tunneling} shows, packets collected outside of smart home (or between \texttt{eth0} and VPN server) are all merged into a single flow (packets with the same destination IP and port~\cite{DBLP:conf/eurosp/TaylorSCM16}), without any original identifier of sender/receiver. To identify the VPN packets and label those with their corresponding devices, we develop a mapping technique based on three observations obtained through our empirical analysis: 
(1) The size of a packet increases after being processed by VPN; (2) Multiple packets with different sizes can have the same size after the encryption performed by VPN; (3) There is a delay of packet transmission caused by VPN, which is usually shorter than 0.02 second.


Observation (1) and (2) can be reasoned through the cryptography algorithms used by \texttt{openvpn} server. It provides three symmetric encryption algorithms -- BF-CBC, AES-128-CBC and DES-EDE3-CBC. All of them are block ciphers through which encryption increases the size of packets. Observation (3) helps us reduce the scope for linking packets before and after VPN. 
As a result, for each VPN-processed packet with timestamp $t$, we first check its \texttt{direction}. If it is inbound, we search its counterpart with smaller packet size in the time window $(t, t+0.02]$. If it is outbound, the time window becomes $[t-0.02, t)$. We measure the effectiveness of our algorithm by counting the rate of successfully pairing and the overall accuracy is 98.8\%.

\ignore{
\begin{enumerate}
	\item Filter all the packets of device $i$ (inbound and outbound) according to the MAC address.
	\item Sort the packets according to the timestamp field.
	\item Extract the traffic window using a sliding window with size $s$ and step $p$.
\end{enumerate}
}

\subsection{Models}
\label{sec:model_select}

In this section, we first describe the baseline model we use for comparison and then describe our customized LSTM-RNN model.

\subsubsection{Baseline Model}
We consider random forest as our baseline model as it has been widely-used in previous works on device fingerprinting~\cite{DBLP:conf/eurosp/TaylorSCM16,DBLP:conf/infocom/SivanathanSGRWV17,Acar2018PeekaBooIS}. A typical random forest is comprised of multiple single decision trees. During the training phase, inner decision trees are trained with different parts of the dataset and a final result is given based on the voting of those separate trees. In our work, we train a random forest model with the labeled packet vectors and the model predicts the device associated with each packet of testing dataset.

Among all the 5 features, \texttt{dport} needs to be processed before being used by the baseline model, since it is a discrete value with wide range (0$\sim$65536) that cannot be directly learned by a machine-learning model.
We first encode \texttt{dport} value into a one-hot binary string.
Since most of the ports are rarely used, we use principal component analysis (PCA) to select 50 principal components from the string. The total variance of them is around 98.9\% according to our statistics, which means the 50 components keep most of  information involved in ports. 

For the hyper-parameters of Random Forest, we set the number of individual trees to 100 to balance training speed and performance. 


\subsubsection{LSTM-RNN Model}

\ignore{
\zl{we have talked about this in previous section}
As the carrier of communication between a device and a remote server, packets generally merge into sequences, which we call traffic traces. To explore the latent influence to classification introduced by neighboring packets, we select LSTM-RNN to model traffic traces. 
}
\ignore{
\begin{figure}[t]
	\centering
	\includegraphics[width=1\columnwidth]{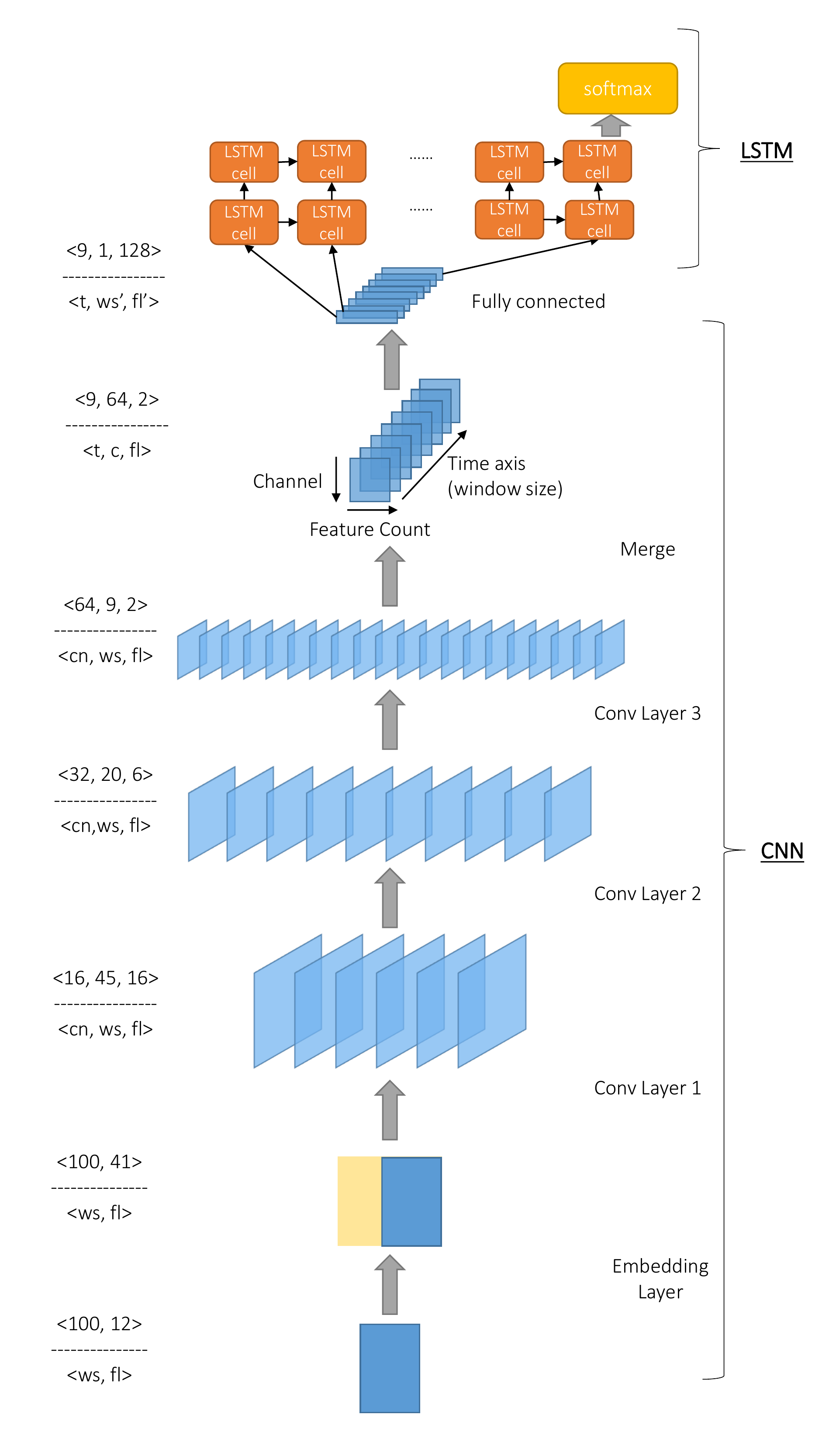}           
	\caption{Structure of CNN-LSTM model (under NAPT setting). $ws$, $fl$, $cn$, $t$ and $c$ are packet window size, feature count, channel number, time axis and channel information.
    }.
	\label{fig:cnnlstm_structure}
	\vspace{-0.2in}
\end{figure}
}

In Section~\ref{sec:lstm_rnn}, we overview the LSTM-RNN and describe its advantage when being used to solve our problem. Below we describe the construction of our LSTM-based models, including a basic version and a bidirectional version. 

\vspace{2pt}\noindent
\textbf{Traffic window.} 
After pre-processing, each packet is transformed into a feature vector. We then group $n$ consecutive vectors to form a \texttt{traffic window}. Figure~\ref{fig:traffic_window} shows an example of a traffic window. With traffic windows, we are able to model the temporal relations of adjacent packets.
We show the impact of window size on evaluation result in Section~\ref{sec:pure_iot_context}.

\vspace{2pt}\noindent
\textbf{Basic LSTM.} A basic version contains multiple blocks and each block contains 4 layers:

\begin{table}[]
\footnotesize
\centering
\begin{tabular}{|c|c|c|c|}
\hline
\textbf{\begin{tabular}[c]{@{}c@{}}Embedding\\ dimension\end{tabular}} & \textbf{\begin{tabular}[c]{@{}c@{}}LSTM hidden\\ dimension\end{tabular}} & \textbf{\begin{tabular}[c]{@{}c@{}}LSTM \\ layer\end{tabular}} & \textbf{\begin{tabular}[c]{@{}c@{}}Dropout \\ rate\end{tabular}}    \\ \hline
30                                                                     & 64                                                                       & 1                                                              & 0.5                                                                 \\ \hline
\textbf{\begin{tabular}[c]{@{}c@{}}Learning\\ rate\end{tabular}}       & \textbf{\begin{tabular}[c]{@{}c@{}}Activation\\ function\end{tabular}}   & \textbf{Optimizer}                                             & \textbf{\begin{tabular}[c]{@{}c@{}}Training \\ epochs\end{tabular}} \\ \hline
0.001                                                                  & ReLU                                                                     & Adam                                                           & 15                                                                  \\ \hline
\end{tabular}
\caption{Parameters of a LSTM-RNN model.}
	\label{tab:basicLSTM}
\end{table}



\noindent -\texttt{Embedding layer.} Embedding has been widely-used in the domain of Natural Language Processing(NLP)~\cite{mikolov2013efficient,sutskever2014sequence} which transforms discrete values into continuous vectors. 
In our LSTM models, we transform \texttt{dport} information into embeddings instead of the one-hot encoding like baseline model, mainly because this embedding layer can be seamless connetcted to other layers. In addition, it is a dynamic ``mini'' neural network gradually updating during the training phase. Therefore, our training data can optimize this representation.

\noindent -\texttt{LSTM layer.} After the processing of embedding layer, the input will be fed into the LSTM layer. At each step, a packet is assigned to a LSTM cell. The output of LSTM cells can be stacked into a matrix as input of the next layer.

\noindent -\texttt{Fully-connected layer.} We put a hidden fully-connected layer between LSTM layer and softmax layer with the size equal to the number of total categories.

\noindent -\texttt{Softmax layer.} The hidden dense layer output is then fed into the softmax layer for normalization. The output of softmax layer is the probability distribution indicating how likely a sample belongs to a category, which sums to one. For our task of multi-class classification, we select the category with the highest probability as the final output.

\begin{figure}[t]
	\centering
	\includegraphics[width=0.85\columnwidth]{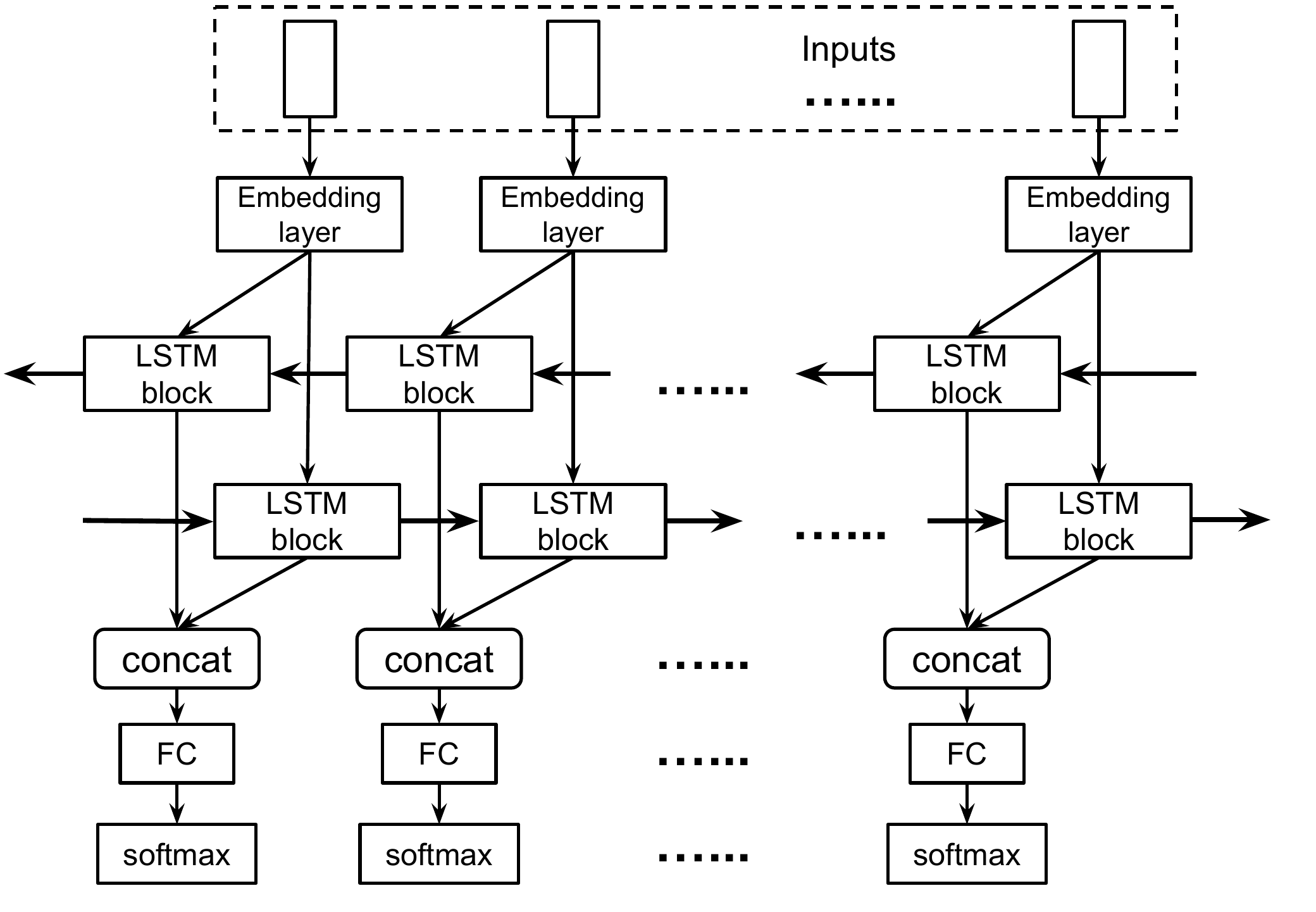}
	\caption{Structure of the bidirectional LSTM used in our work.}
	\label{fig:blstm}
	\vspace{-0.2in}
\end{figure}

\vspace{2pt}\noindent
\textbf{Bidirectional LSTM.} The basic LSTM model only looks into the ``past'' of a packet when learning contextual information. Bidirectional LSTM (BLSTM) is an extension to the basic LSTM, which utilizes the information from ``future'', by combining another LSTM layer moving from the end of a sequence to its beginning~\cite{Goodfellow-et-al-2016}. In areas like phoneme classification~\cite{Graves2005FramewisePC} and sequence tagging~\cite{Huang2015BidirectionalLM}, BLSTM significantly improves the performance compared to a traditional one.
Since our model works on a traffic window which consists of multiple adjacent packets, we can utilize the information of the packets transmitted \textit{after} the current packet to classify it. 
The main change we apply on the LSTM layer is to concatenate cell states of backward and forward LSTM layers and feed them to the dense layer. 
Figure~\ref{fig:blstm} shows the structure of the bidirectional LSTM used in our work.


\ignore{
\subsection{Device Identification}
\label{sec:system_device_identification}

\begin{table*}[tbp]
	\centering
	\caption{Details of feature vectors. }
	\label{tab:fs}
	\begin{tabular}{|l|l|l|}
	\hline
	Acronym & Features & Experiments  \\
	\hline
    \hline
	$FS_N$ & $<$\texttt{dport, protocol, direction, packetsize, timeinterval}$>$  & Pure-traffic, Mix-traffic (NAPT)\\
	\hline
	$FS_{NV}$ & $<$\texttt{direction, packetsize, timeinterval}$>$  & Pure-traffic, Mix-traffic (NAPT+VPN)\\
	\hline
    $FS_{WL}$ &  $<$\texttt{subtype, framesize, timeinterval}$>$ & Wireless \\
	\hline
	\end{tabular}
\end{table*}


The first goal of \system\ is to accurately identify the IoT devices installed in a smart home environment. We consider this is a multi-class classification problem and the trained DNN predicts the device model (e.g., Samsung SmartCam) using a group of packets as input.
As described in Section~\ref{sec:adversary}, we assume the adversary only has access to the encrypted traffic. In addition, the device identifier, e.g., MAC or the original source IP address, is not always available. Such network environment is more complicated comparing to previous works~\cite{DBLP:conf/infocom/SivanathanSGRWV17, DBLP:journals/corr/abs-1708-05044}, in which the flows of different devices are isolated. Below we elaborate how the feature vector and DNN are customized for different network environment.



\vspace{2pt} \noindent
\textbf{Classification of packet window.} We consider three environment:  \textit{NAPT}, \textit{NAPT+VPN} and \textit{wireless} environment. In particular, some features are removed or modified from $FS$, due to the restriction of adversary's visibility into the traffic. Table~\ref{tab:fs} summarizes the feature vector for each scenario.
For all those scenarios, DNN takes in a packet window and outputs the device model with the highest probability after classification, if the probability is over a pre-defined threshold. 

\subsubsection{NAPT-enabled environment.} 
\label{sec:napt_env}
In this scenario, we assume the Ethernet adversary has access to the packets between a gateway and remote service. NAPT is enabled on the gateway, so none of the device identifiers are preserved (MAC and source IP addressed are replaced by gateway's), as described in Section~\ref{sec:smartnetwork}. We name feature vector under this scenario as $FS_N$, which is the same as $FS$.

\ignore{
As Figure ~\ref{fig:adversary} shows, devices in a smart home environment usually connect to the Internet through a NAPT-enabled router.\zl{we need evidence for ``usually''} NAPT\cite{rfc2663} is a kind of network address translation technique which allows devices within a subnet to share the same public IP address of the gateway and to communicate with the Internet. For outgoing packets of a subnet, the source IP address and the source port number are modified by the router. Similarly, the destination IP address and the destination port number are modified for incoming packets. As such, we cannot differentiate devices based on $sip$ and $sport$ from the outside of gateway because packets from a device may be mapped to multiple ports of the gateway and the ports could be changed dynamically from time to time. Apart from that,  for a packet generated by a subnet device, its transmitter address in the frame header will be overwritten to that of the router, which means the MAC address of a packet cannot be used to differentiate its source.

For the traffic in an NAPT-enabled environment, we conduct the data preprocessing procedure shown in Figure ~\ref{fig:data_preprocess}. One step of data preprocessing is to discard three features $<sip, dip, sport>$ from traffic windows. The reason why removing $dip$ feature from the dataset is based on the following consideration: The pervasion of cloud services, like Amazon Web Service, Google Cloud, has made more and more corporations migrate their back-end servers to the cloud. However, the IP address of a cloud virtual machine is usually fluctuating for it is randomly picked from the public IP pools maintained by the cloud service provider. Therefore, even the destination IP address may rise the classification performance to some extent, we assume it is not helpful to the generalization of the model. After the removal, an input matrix of size $(m,5)$ is generated, where $m$ is the traffic window size and $5$ is the number of remaining features. 
}

Different from other 4 features in $FS$, the feature value of \texttt{dport} is quite sparse, ranging from 0 to a very large positive number. DNN is known to perform poorly on sparse feature~\cite{embedding}, so we map a \texttt{dport} feature value into a much dense range, by introducing another embedding layer \texttt{Embed(x)} (details of embedding layer are described in Section~\ref{sec:neural_structure}). 
In particular, we set the output dimension $x$ of the embedding layer to 30, meaning that a port number will be mapped to an element in a vector with 30 elements.
The embedding layer will be continuously updated when new input comes to DNN. After embedding, $FS_N$ is expanded to a vector of 41 elements.



\subsubsection{NAPT- and VPN-enabled environment.}
We also consider the case when both NAPT and VPN are enabled by the gateway. 
Prior work~\cite{DBLP:journals/corr/ApthorpeRF17} also argued enabling VPN is not sufficient against network eavesdroppers (in Section 6.2 ``Tunneling traffic'' of that paper), but traffic-analysis method or evaluation is not given in this case. We attempt to verify this argument with concrete evidence.
We assume the adversary has access to the network link between gateway and the VPN exit. Comparing to the NAPT-enabled environment, not all the features in $FS$ can be used.
VPN software encrypts each original packet into ciphertext and wraps it into a new IP packet with its payload being the ciphertext. The new packets will be transmitted from the gateway to VPN exit where it is  decrypted. As such, the packets intercepted by the Ethernet adversary are all \texttt{IP-TCP} packets. As such, both \texttt{dip} and \texttt{dport} are not meaningless to the adversary. In addition, the original \texttt{protocol} is obscured by the VPN software. 
Therefore, we remove \texttt{dport} and \texttt{protocol} from the packet window, leaving only 3 features ($<$\texttt{direction, packetsize, timeinterval}$>$). This new feature vector is named $FS_{NV}$.

Consequently, the new feature vector $FS_{NV}$ is much ``slimmer'' comparing to $FS_{N}$ of the NAPT scenario, as the dropped \texttt{dport} and \texttt{protocol} can be converted to 30 and 7 elements after embedding and one-hot encoding. Still, we found such issue can be addressed by modifying the size of the convolutional kernels in DNN. Table~\ref{tab:network_structure} describes the details of each layer under this setting (row 4 to 6).

\subsubsection{Wireless environment.} 
Under this scenario, the adversary has access to the packets transmitted through Wi-Fi and we assume the communication is properly protected by protocols like WPA and WPA2. As the encryption happens at the data-link layer, none of the information at higher layers like network layer and transport layer can be obtained.

As described in section ~\ref{sec:pre-process}, we extract the data frames after pre-processing. As a result, each packet group consists of a series of data frames, and the feature vector is in fact $<$\texttt{subtype, framesize, timeinterval}$>$, which maps to $<$\texttt{protocol, packetsize, timeinterval}$>$ of the raw representation. We name this feature vector as $FS_{WL}$.
The 802.11 standard defines 15 subtypes for the data frames~\cite{dataframe}. We use a one-hot encoding to represent \texttt{subtype} field, similar to the \texttt{protocol} field. In the end, $FS_{WL}$ is converted into a vector with 17 elements.

\vspace{2pt} \noindent
\textbf{Unused features.} 
Some features leveraged by previous works~\cite{DBLP:journals/corr/ApthorpeRF17,DBLP:conf/infocom/SivanathanSGRWV17} are not utilized by \system. Since the packet window only covers a short period of time, we discard the features requiring long observation window, e.g., the interval between NTP packets and the sleeping time of a device. We assume the DNS resolution result could be cached by the IoT device, hub or gateway for a long time, so domain names are not used by \system\ as well.
On the other hand, our DNN model is able to learn the optimal feature representation  automatically and achieves even better effectiveness, as shown in Section~\ref{sec:evaluation}.


\subsection{Activity Inference}
\label{sec:activity}
The majority of IoT devices are designed to directly interact with human or the surrounding physical environment. When a pre-defined logic condition is satisfied, like the door protected by SmartLock is unlocked, a notification will be sent to service provider or user through network. The important question we want to answer here is whether it is sufficient to infer private information about user's activity simply from the captured network traffic.

The answer seems to be positive from our exploratory analysis. Below we give a few examples.

\begin{itemize}
\item \textbf{Google voice assistant.} The voice assistant we experimented stay in the standby mode when not interacted with human. Once it receives a voice command (e.g., \textit{Hey google, play music}), it will immediately turn into the working mode. In particular, a request will be sent to the service provider and the response carries the computation results will be sent back. As a result, a traffic burst will be observed. The traffic pattern is illustrated in Figure~\ref{fig:google_voice}.  

\item \textbf{Orvibo switch.} An Orvibo switch can be controlled through a paired mobile app. According to our observation, the switch continually sends a 192-byte heartbeat packet to the remote server every 30 seconds and receives a response packet with the size of 176 bytes. When the user clicks the control button on the mobile app UI, a series of packets with the size of 448, 224, 240, 272 bytes respectively will be transmitted, as shown in Figure ~\ref{fig:orvibo_switch}.

\item \textbf{Xiaomi camera.} Xiaomi camera provides two modes to transmit the recorded scenes to a user -- WLAN and WAN. When the user uses the same gateway as the camera, the real-time scenes will be transmitted to the user's mobile phone through WLAN (e.g., Wi-Fi). 
When the user is beyond the range, the scenes will be transmitted to the remote server in \texttt{UDP} protocol. The user can use the mobile app to fetch the scenes from the server. As such, the traffic volume in the later scenario is much larger than that in the prior scenario, when measured by an Ethernet adversary: according to our experiment, the traffic rate of WLAN mode is 0.237KB/s in average, whereas it is 120.2KB/s for WAN mode, as shown in Figure~\ref{fig:xiaomi_camera}. The big difference of traffic rate can be utilized by the adversary to infer whether the host is at home or not. 
\end{itemize}



If the unique fingerprint corresponding to traffic burst or rate can be identified, the user's activities or location can be inferred.
We reuse the DNN models for device identification this scenario but using different training dataset. The output of DNN model is a user behavior related to one IoT device model (e.g., turn on Belkin Wemo Switch) inferred from a single packet window.
During evaluation, we set up an environment with 8 IoT devices and simulate 19 user behaviors that can be handled by them. The list of behaviors is shown in Table ~\ref{tab:device_activity}. Below we elaborate how the user behaviors are simulated.


\vspace{2pt}\noindent
\textbf{User-behavior simulation.} 
The 10 tested IoT devices can be classified into three types, \textit{Smart assistants}, \textit{Hubs} and \textit{Smart Accessories}. Except smart assistants which are controlled by voice, all the other
devices can be controlled by a mobile application (app) directly or indirectly if connected to an IoT hub. The communication topology of IoT devices in our experiment is depicted in Figure~\ref{fig:dev_topology}.
Figure~\ref{fig:side:a} shows the mobile app paired with Xiaomi Hub. 
All user's IoT devices of the same vendor, like Xiaomi camera, can be controlled by this app.
In addition to vendor's apps, third-party apps following IFTTT standard~\cite{url_ifttt} can be leveraged to control devices as well. 
Figure ~\ref{fig:side:b} shows a setting panel of an IFTTT-based app. 
We experiment with both vendor apps and IFTTT apps to interact with the devices. 


\begin{figure}[t]
	\centering
	\includegraphics[width=0.70\columnwidth]{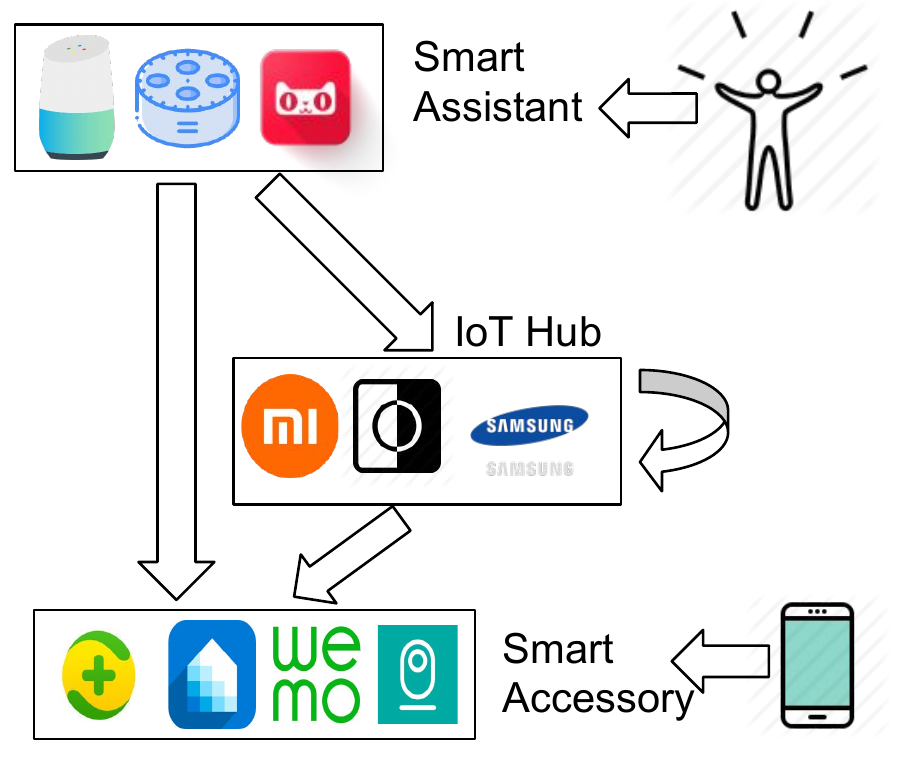}         
	\caption{Device communication topology.}
	\label{fig:dev_topology}
\vspace{-0.2in}
\end{figure}

\begin{figure}
	\begin{minipage}[t]{0.5\columnwidth}
		\centering
		\includegraphics[width=1.75in]{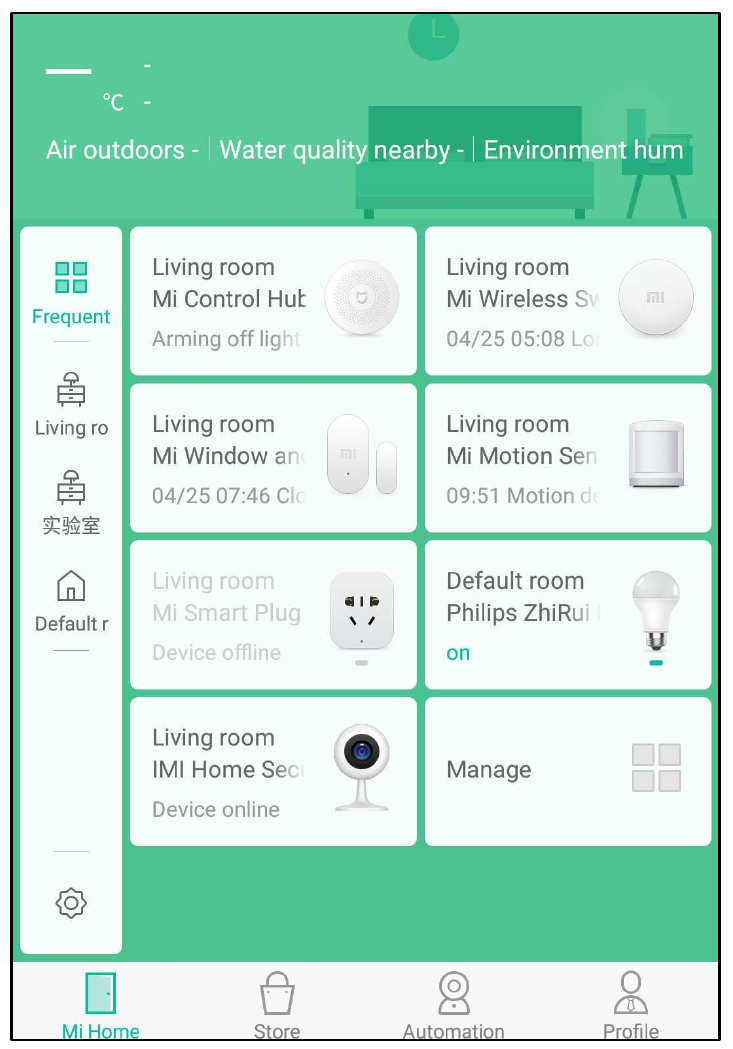}
		\caption{Mi Home App}
		\label{fig:side:a}
	\end{minipage}%
	\begin{minipage}[t]{0.5\columnwidth}
		\centering
		\includegraphics[width=1.75in]{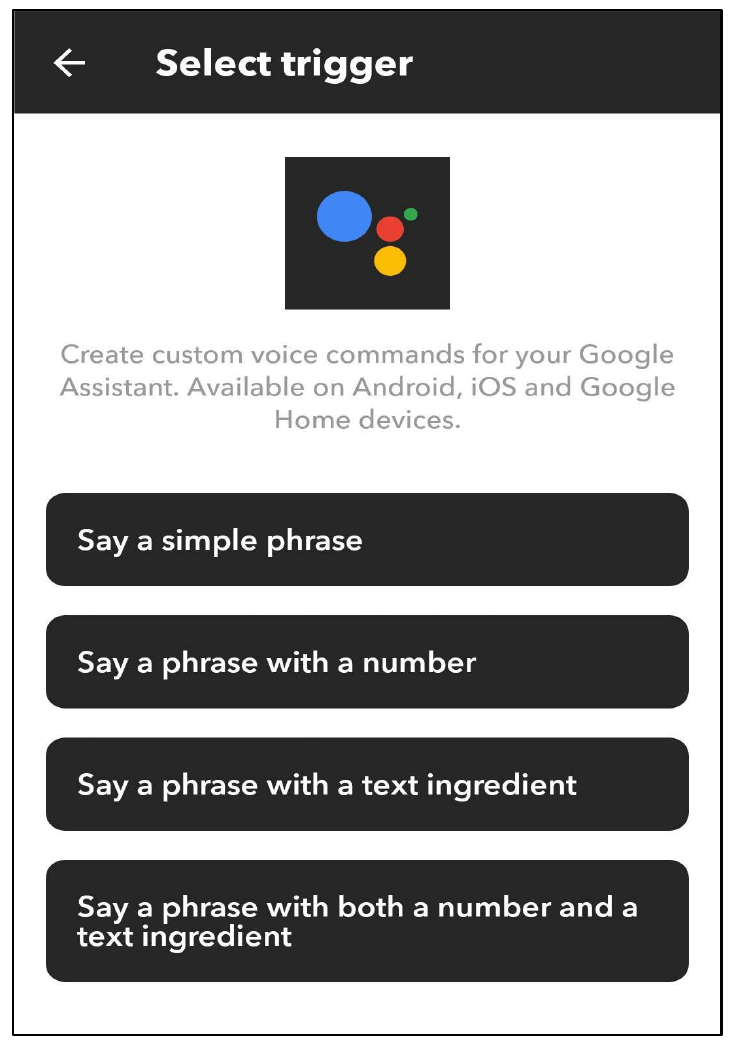}
		\caption{IFTTT App}
		\label{fig:side:b}
	\end{minipage}
\vspace{-0.2in}
\end{figure}


For each activity, we need an adequate amount of samples to train DNN. Asking human to generate the sheer amount of samples is quite labor-intensive. So we automate the process using the methods below.

\begin{enumerate}
	\item \textit{Voice-command generation.} We record a list of voice commands and play each command for 200 times when a voice assistant is to be trained or tested.
	\item \textit{App-action generation.} We use MonkeyRunner provided by Google~\cite{url_monkeyrunner} to automatically inject control actions to the app UI. Each action is repeated for 200 times and a short interval is kept between two actions. 
\end{enumerate}


\begin{table}[t]
	\centering
	\caption{IoT device activities}
	\label{tab:device_activity}
	\scalebox{0.92}[0.92]{%
		\begin{tabular}{|l|l|l|l|}
			\hline
			\textbf{Device} & \textbf{Type} &\textbf{Activity} & \textbf{Trigger Action}                                                                            \\
			\hline
            \hline
			\multirow{4}*{Amazon Echo Dot} & \multirow{4}*{Smart assistant} &
			Standby &
			\multirow{4}*{Voice command} \\ \cline{3-3}
			~& & Control Philips bulb &\\
			\cline{3-3}
			~& & Control Wemo switch &\\
			\cline{3-3}
			~& & Control TP-Link switch &\\
	
			\hline
			\multirow{4}*{Tmall Assistant} &\multirow{3}*{Smart assistant} &Standby & \multirow{4}*{Voice command}\\
			\cline{3-3}
			~ & & Wakeup &\\
			\cline{3-3}
			~ & &Control Tospo bulb &\\
			\cline{3-3}
			~ & &Play song & \\
			\hline
			\multirow{3}*{Google Home}&\multirow{3}*{Smart assistant} & Standby & \multirow{3}*{Voice command}\\
			\cline{3-3}
			~ & &Wakeup & \\
			\cline{3-3}
			~ & &Ask question &\\
			\hline
			
			Samsung SmartThings & Hub & Control Wemo Switch & App action \\			
			\hline
			\multirow{3}*{Xiaomi Hub} & \multirow{3}*{Hub} & Control Xiaomi bulb & \multirow{3}*{App action}\\
			\cline{3-3}
			~ & & Live monitor(WLAN) &\\
			\cline{3-3}
			~ & & Live monitor(WAN) &\\
			\hline
			\multirow{2}*{Belkin Wemo Switch} & \multirow{2}*{Accessory} & Standby & \multirow{2}*{App action}\\		
			\cline{3-3}
			~ & & Turn on/off & \\
			\hline
	
			Tmall Box & Accessory & Play movies & App action\\
			\hline	
			\multirow{2}*{Orvibo Switch} &\multirow{2}*{Accessory} &Standby &\multirow{2}*{App action}\\
            \cline{3-3}
			~& & turn on/off & \\
			\cline{3-3}
			
			\hline
		\end{tabular}
	}
\end{table}
\vspace{-0.05in}
}

	\section{Evaluation}
\label{sec:evaluation}

In the evaluation, we want to understand how our models perform under different network scenarios. In this section, we first introduce the datasets we used for evaluation and our evaluation metrics. Then we describe our three scenes and the corresponding results. Finally we show several case studies.

As a quick overview of our results, we found LSTM-RNN models can well handle packet identification tasks with an overall accuracy over 92.0\% in NAPT and VPN configurations on IoT traffic. Compared to basic LSTM, bidirectional LSTM performs better, suggesting the packet dependency indeed reveals the patterns unique to each individual IoT device.
\subsection{Experiment Settings}
\label{sec:eval_method}

\noindent \textbf{Scenarios.} We evaluate \system\ in two scenarios -- \texttt{pure-IoT} (only one active IoT device) and \texttt{noisy} (multiple active IoT and non-IoT devices). In each scenario, we evaluate \system\ with two different gateway configurations -- \texttt{NAPT} and \texttt{VPN}. 

\noindent \textbf{Datasets.} We constructed two datasets for the two scenarios. Each dataset has an NAPT version and a VPN version. We split each dataset with the training and testing ratio of 8:2 and conduct 5-fold cross-validation on it. 
Below are the details of each dataset.

\begin{figure}[t]
	\centering
	\includegraphics[width=1.0\columnwidth]{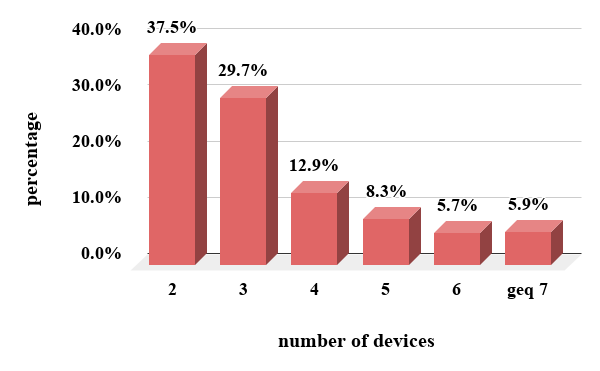} 
	\caption{Proportion of device combinations. (geq: greater than or equal to 7)}
	\label{fig:num_devices}
	\vspace{-0.2in}
\end{figure}

\begin{enumerate}
    \item \texttt{Dataset-Ind.} This dataset contains traffic representations from 10 individual IoT devices. To facilitate the training process of LSTM-RNN models, the dataset is organized into collections of traffic windows, each traffic window only contains packets from \texttt{one certain device}. To make the dataset more balanced, we set a number threshold 5,000 for each device. All IoT devices own 5,000 randomly-selected samples except Xiaomi hub, tplink plug, orvibo plug and broadlink plug, due to that they generate much fewer packets than others. In total, \texttt{Dataset-Ind} contains 32,760 traffic windows.
    \item \texttt{Dataset-Noise.} This dataset is collected by keeping multiple devices active in the same time period. As a result, the traffic windows in this dataset are composed of packets from \textit{more than} one devices. In total, \texttt{Dataset-Noise} includes 114,989 traffic windows. Figure~\ref{fig:num_devices} shows the distribution of device combinations. From it we can see, 2-device and 3-device combinations are most common.
    
\end{enumerate}

\ignore{
\begin{figure}[t]
	\centering
	\includegraphics[width=1.0\columnwidth]{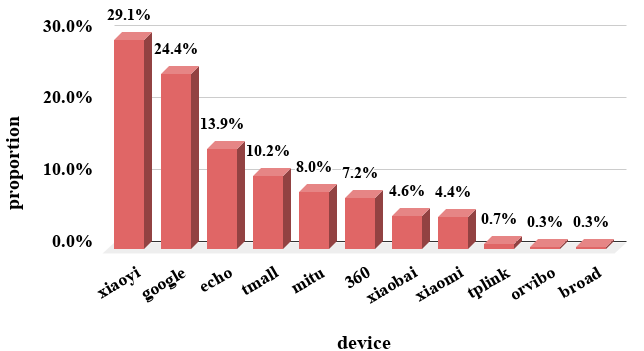} 
	\caption{Proportion of packets in Dataset-Ind.}
	\label{fig:dataset_ind}
	\vspace{-0.2in}
\end{figure}

\begin{figure}[t]
	\centering
	\includegraphics[width=1.0\columnwidth]{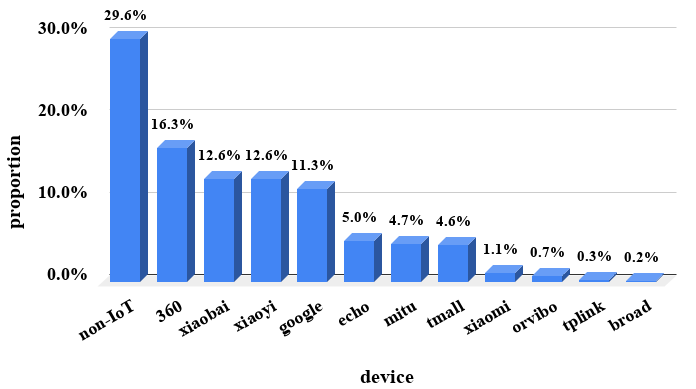} 
	\caption{Proportion of packets in Dataset-Noise.}
	\label{fig:dataset_noise}
	\vspace{-0.2in}
\end{figure}
}

\noindent \textbf{Metrics.} Since \system\ is able to classify individual packets, we measure the effectiveness of \system\ based on the probability that the device is correctly identified \textit{per packet}. We use \textit{overall accuracy} (similar to~\cite{DBLP:conf/ccs/SirinamIJW18}) and \textit{category accuracy} for our case. For overall accuracy, we count $N$ as all the packets and $P_{correct}$ as the total number of correctly classified packets, and compute $\frac{P_{correct}}{N}$. For category accuracy, we assess how \system\ performs on each device. For device $A$, we count $NA$ as all packets belonging to $A$ and $PA_{correct}$ as $A$'s packets correctly classified under $A$. The category accuracy for $A$ is $\frac{PA_{correct}}{NA}$. As an example, the diagonal cells on the confusion matrix shown in Figure~\ref{fig:cm_rf_noisy_napt} describe the category accuracy.

\ignore{
To measure the effectiveness of our models, we use two metrics. 1) Accuracy, which equals to 
$ \frac{TP+TN}{TP+FP+TN+FN}$, where $TP, TN, FP, FN$ 
refer to true positive, true negative, false positive and false negative correspondingly. 
2) Confusion matrix. For multi-class classification tasks, confusion matrix is a good option to illustrate the performance of a model. Figure~\ref{fig:cm_rf_noisy_napt} shows an example of confusion matrix. Elements on the diagonal are correctly-classified samples and others are mis-classified. A confusion matrix not only indicates whether a category is well recognized by the model but illustrates which categories cannot be clearly distinguished by the model.
}

\subsection{Pure-IoT Scenario}
\label{sec:pure_iot}

In this setting, there is only one active IoT device working during a period of time. Therefore, we assess how \system\ performs when traffic is not merged. In practice, such scenario happens when the rest IoT devices enter hibernation mode.

\subsubsection{Baseline Model}
\label{sec:single_pkt_class}

\begin{table*}[]
\footnotesize
\centering
\scalebox{1}[1]{%
\begin{tabular}{|c|c|c|c|c|c|c|c|c|c|c|c|c|}
\hline
\textbf{config}       & \textbf{model} & \textbf{average} & \textbf{\begin{tabular}[c]{@{}c@{}}echo \\ dot\end{tabular}} & \textbf{\begin{tabular}[c]{@{}c@{}}google \\ home\end{tabular}} & \textbf{\begin{tabular}[c]{@{}c@{}}tmall\\ assistant\end{tabular}} & \textbf{\begin{tabular}[c]{@{}c@{}}xiaomi\\ hub\end{tabular}} & \textbf{\begin{tabular}[c]{@{}c@{}}360\\ cam\end{tabular}} & \textbf{\begin{tabular}[c]{@{}c@{}}tplink\\ plug\end{tabular}} & \textbf{\begin{tabular}[c]{@{}c@{}}orvibo\\ plug\end{tabular}} & \textbf{\begin{tabular}[c]{@{}c@{}}mitu\\ story\end{tabular}} & \textbf{\begin{tabular}[c]{@{}c@{}}xiaobai\\ camera\end{tabular}} & \textbf{\begin{tabular}[c]{@{}c@{}}broadlink \\ plug\end{tabular}} \\ \hline
\multirow{3}{*}{NAPT} & RF             & 92.2             & 89.0                                                         & 85.9                                                            & 86.9                                                               & 89.6                                                          & 99.0                                                       & \textbf{99.9}                                                  & \textbf{99.9}                                                  & 93.3                                                          & 98.5                                                              & 99.3                                                               \\ \cline{2-13} 
                      & LSTM           & 97.3             & \textbf{98.5}                                                & 91.6                                                            & 93.9                                                               & 98.6                                                          & \textbf{99.9}                                              & \textbf{99.9}                                                  & \textbf{99.9}                                                  & 98.7                                                          & \textbf{99.9}                                                     & \textbf{99.9}                                                      \\ \cline{2-13} 
                      & BLSTM          & \textbf{99.2}    & 97.0                                                         & \textbf{99.2}                                                   & \textbf{99.8}                                                      & \textbf{99.9}                                                 & \textbf{99.9}                                              & \textbf{99.9}                                                  & \textbf{99.9}                                                  & \textbf{99.3}                                                 & \textbf{99.9}                                                     & \textbf{99.9}                                                      \\ \hline
\multirow{3}{*}{VPN}  & RF             & 83.2             & 76.1                                                         & 81.2                                                            & 74.7                                                               & 94.0                                                          & 83.2                                                       & 89.1                                                           & 93.1                                                           & 87.5                                                          & 90.5                                                              & \textbf{99.0}                                                      \\ \cline{2-13} 
                      & LSTM           & 92.4             & 89.7                                                         & 89.7                                                            & 75.4                                                               & 96.1                                                          & 95.9                                                       & 92.2                                                           & 95.5                                                           & 96.8                                                          & 94.7                                                              & 95.7                                                               \\ \cline{2-13} 
                      & BLSTM          & \textbf{97.7}    & \textbf{96.6}                                                & \textbf{96.8}                                                   & \textbf{94.7}                                                      & \textbf{99.4}                                                 & \textbf{98.5}                                              & \textbf{98.0}                                                  & \textbf{99.5}                                                  & \textbf{98.9}                                                 & \textbf{99.7}                                                     & 96.7                                                               \\ \hline
\end{tabular}
}
\caption{Accuracy of baseline model under pure-IoT scenario \\ (\textbf{RF}, \textbf{LSTM} and \textbf{BLSTM} stand for random forest, basic LSTM and bidirectional LSTM respectively).}
	\label{tab:acc_pureIoT}
\end{table*}

We first evaluate the performance of our baseline model, Random Forest. In this scenario, 
The purposes are two-fold: (1) To explore the feasibility of classifying individual packet without context; (2) To evaluate the effectiveness of features we selected from packet's metadata.


\noindent \textbf{Experimental results.} Table~\ref{tab:acc_pureIoT} shows the accuracy of random forest in NAPT and VPN configuration. As can be seen, with NAPT configuration, the random forest can reach a high identification accuracy on most IoT devices. Among them, smart plugs and network cameras have the highest accuracy while voice assistants have a lower accuracy $\sim87\%$. 
Compared with NAPT, random forest performs worse in VPN configuration with a 9.0\% decline in overall accuracy. Accuracy on voice assistant is affected most.

\noindent \textbf{Result analysis.} We first use the built-in API provided by \texttt{scikit- learn} library to obtain feature importance. The results show that in NAPT,  \texttt{dport}, \texttt{frame length}, \texttt{time interval} and \texttt{protocol} hold an importance factor of 55.5\%, 22.8\%, 12.8\% and 8.0\% separately. In VPN, \texttt{frame length} and \texttt{time interval} take up around 54.9\% and 43.0\% separately.

Compared with NAPT configuration, the obvious decline in VPN mainly comes from the change of \texttt{dport} and \texttt{protocol} (\texttt{dport} and \texttt{protocol} information is not preserved in the packet between gateway and VPN server) and partial loss of \texttt{frame length} due to  padding by VPN client.



\subsubsection{LSTM-RNN Models}
\label{sec:pure_iot_context}

We conjecture that the dependency of packets can be used for device identification, and we model it through LSTM-RNN models. Below we evaluate the two proposed LSTM-RNN models on \texttt{Dataset-Ind}. 



\noindent \textbf{Experimental results.} Table~\ref{tab:acc_pureIoT} also shows performance of \texttt{basic LSTM} and \texttt{bidirectional LSTM} when the input traffic window contains 100 consecutive packets. 
We can see that compared to the baseline, both of the models have seen increase of accuracy on most devices. 
The result also shows LSTM-RNN models can well handle IoT devices producing large volume of traffic like voice assistant.

\begin{figure}[t]
	\centering
	\includegraphics[width=1.0\columnwidth]{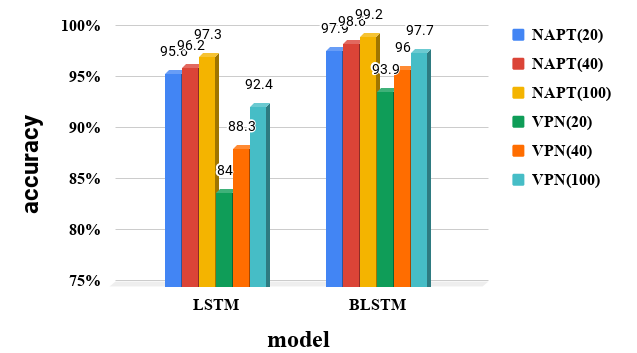} 
	\caption{The impact of traffic window size (pure-IoT).}
	\label{fig:window_length}
	\vspace{-0.2in}
\end{figure}

\noindent \textbf{Impact of traffic window size.} 
We compare the accuracy of LSTM-RNN models with different window sizes: 20, 40 and 100. The result is shown in Figure~\ref{fig:window_length}. From it we can see for both NAPT and VPN configurations, LSTM-RNN models perform better when the traffic window size grows. This result indicates the relation between packets with long timing gap can still provide useful information for our models.
In the following sections, we take 100 as the default size of our traffic window. 

\subsection{Noisy Scenario}

In this section, we evaluate the impact of non-IoT device traffic on our task. 
\subsubsection{Baseline Model}
We first tested Random Forest using \texttt{Dataset- Noise} to understand the impact of non-IoT traffic and traffic fusion.
Given the different traffic volume among devices, the packets we collect are imbalanced, as Figure~\ref{fig:dataset_statis} shows.

\noindent \textbf{Experimental results.} 
Compared with pure-IoT scenario, Random Forest has a prominent decline in the overall accuracy, reaching 84.5\% in NAPT and 67.6\% in VPN.
We also use confusion matrix across devices to show the classification results by category with non-IoT traffic in NAPT and VPN configuration (see Figure~\ref{fig:cm_rf_noisy_napt} and Figure~\ref{fig:cm_rf_noisy_vpn} in Appendix~\ref{sec:appendix_performance_pureIoT}). 
From them we can see, \texttt{voice assistants}, like Echo Dot, Google Home and Tmall Assistant, see larger performance drop compared to other IoTs, with a $\sim25\%$ decline in NAPT and a $\sim50\%$ decline in VPN configurations. 

\noindent \textbf{Result analysis.} We manually check the mis-classified packets and find that most of them are transmitted through ports 443 and 80. The primary reason is that those ports are likely to be used by different IoT and non-Iot at the same time, so our model is more likely to be confused.

\subsubsection{LSTM-RNN Models}

\begin{figure}[t]
	\centering
	\includegraphics[width=1.0\columnwidth]{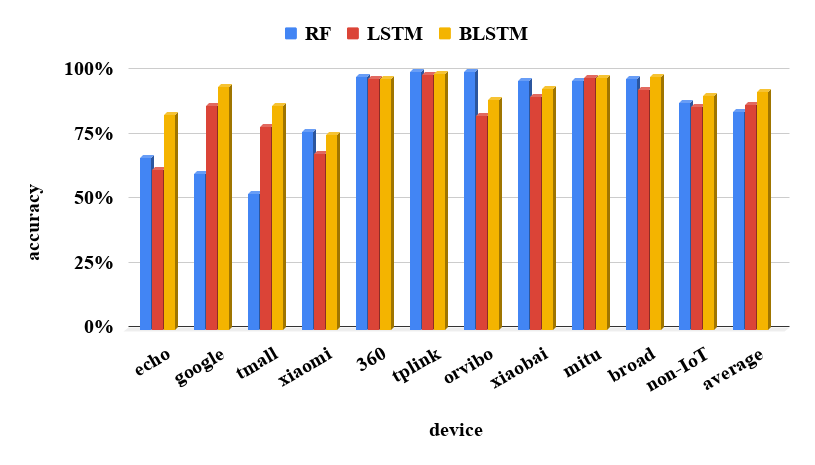} 
	\caption{Performances of three models in NAPT environment.}
	\label{fig:compare_three}
	\vspace{-0.2in}
\end{figure}

\begin{figure}[t]
	\centering
	\includegraphics[width=1.0\columnwidth]{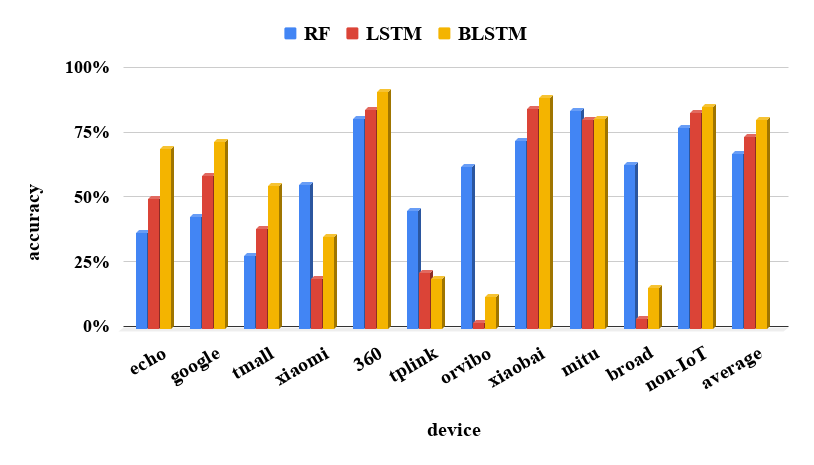} 
	\caption{Performances of three models in VPN environment.}
	\label{fig:compare_three_vpn}
	\vspace{-0.2in}
\end{figure}

\noindent \textbf{Experimental results.} Figure~\ref{fig:compare_three} and Figure~\ref{fig:compare_three_vpn} give the comparison results between three models in NAPT and VPN configurations. The last group of columns (``average'') shows the overall accuracy. 
Bidirectional LSTM achieves the highest accuracy of 92.1\% in NAPT and 81.0\% in VPN. Basic LSTM reaches 87.1\% and 74.1\%.
Figure~\ref{fig:cm_blstm_noisy_napt} and Figure~\ref{fig:cm_blstm_noisy_vpn}  of Appendix~\ref{sec:traffic_pattern} show the concrete classification results of bidirectional LSTM by device categories in two configurations. 
From them we know: (1) LSTM-RNN models are good at recognizing traffic in NAPT configuration; 
(2) LSTM-RNN performs much worse in VPN configuration, especially on IoT devices like smart plugs (BLSTM: 12.6\% for Orvibo, 20.4\% for Tplink and 15.9\% for Broadlink in 5 cross validations).

\noindent \textbf{Result analysis.} 
From the above results we can see LSTM-RNN models fail to classify traffic generated by smart plugs in VPN configuration. We find this observation can be ascribed to the sparse traffic generated by the devices. Due to the relative long time intervals between packets and the low packet amount, the traffic generated by smart plugs can be easily ``overwhelmed'' by traffic from others, leading to the original relation between packets being impaired. In extreme cases, packets generated by smart plugs can be ``diluted'' to less than 3\% in a traffic window (3 in 100 packets). 
The situation becomes worse when distinctive features like \texttt{dport} and \texttt{frame length} are more likely to be confused. 
In Section~\ref{sec:case_study} we show a case of \texttt{orvibo plug}.

\ignore{
\noindent
\textbf{Evaluation metrics.} 
In our evaluation, we use three metrics to measure the performance of \system\ -- accuracy, precision and recall. To notice, the label is assigned to individual packet window. $TP$, $TN$, $FP$ and $FN$ refer to the number of true positives, true negatives, false positives and false negatives.

\begin{itemize}
	\item \textbf{Accuracy} is calculated by the formula $Acc=\frac{TP+TN}{TP+FP+FN+TN}$.
    In our case, accuracy measures the ratio of packet windows that are correctly classified.
	\item \textbf{Precision} is calculated by $Prec = \frac{TP}{TP+FP}$. The precision of a model depicts its ability of avoiding labeling a negative sample as positive.
	\item \textbf{Recall} is calculated by $Rec = \frac{TP}{TP+FN}$. The recall demonstrates the model's ability of avoiding labeling a positive sample as negative.
\end{itemize}

Though we could compute the three metrics on all packets across devices, the result would be biased when one device generates far more packets than others. As such, we first compute the metrics per device, and then aggregate the metrics using weighted average. As example, assuming the number of packets for device $A$ and $B$ are 400 and 600, and the accuracies are 0.8 and 0.6. The weight will be 0.4 ($\frac{400}{400+600}$) and 0.6, and the overall accuracy will be  0.68 ($0.4 * 0.8 + 0.6 * 0.6$).


\begin{table}[t]
	\centering
	\caption{Devices used in wireless sniffing attack}
	\label{tab:device_list}
	\begin{tabular}{|l|l|l|}
		\hline
		\textbf{Device} & \textbf{Type} & \textbf{IoT or not} \\
        \hline
		Google Home & Smart assistant & IoT \\
		\hline
		Amazon Echo Dot & Smart assistant & IoT \\
		\hline
		Tmall Assistant & Smart assistant & IoT \\
		\hline
		Xiaobai camera & Accessory & IoT\\
		\hline
		Kongke Switch & Accessory & IoT \\
		\hline
		Sony Phone	& Mobile phone & non-IoT \\
		\hline
		Motorola Phone & Mobile phone & non-IoT \\
		\hline
		Netcore Router &  Gateway & non-IoT \\
		\hline
	\end{tabular}
\end{table}


\begin{table*}[t]
	\centering
	\caption{Experiments on the datasets. }
	\label{tab:evaluate}
	\begin{tabular}{|l|l|l|l|}
	\hline
	Test & \texttt{Dataset-Pub} & \texttt{Dataset-Act} & \texttt{Dataset-WL} \\
	\hline
    \hline
    Adversary & Ethernet  & Ethernet & Wireless \\
	\hline
	Device Identification & \system\ and \texttt{AppScanner}  & \xmark & \system \\
	\hline
	Activity Inference & \xmark & \system\ & \xmark \\
	\hline
	\end{tabular}
\end{table*}

}




\ignore{
\subsection{Device Identification}
\label{sec:eval_device}

We used the \texttt{Dataset-Pub} to evaluate how \system\ performs when Ethernet packets are available. In particular, we compare \system\ to an existing work named \texttt{AppScanner}~\cite{DBLP:conf/eurosp/TaylorSCM16} for a ``pure'' traffic setting and a ``mixed'' traffic setting. Finally, we evaluate the effectiveness of \system\ using \texttt{Dataset-WL}.




\subsubsection{AppScanner}

We consider \texttt{AppScanner}~\cite{DBLP:conf/eurosp/TaylorSCM16}, a work published recently aiming at fingerprinting mobile apps from encrypted network traffic, as the target for performance comparison. The main reason of choosing \texttt{AppScanner} is that it works on the same network protocols (Ethernet) and the raw input representation is similar. Because the system is not open-sourced, we reimplemented the system following the description of the paper.



Different from \system\ which leverages DNN for feature representation learning, \texttt{AppScanner} selects features through a manual feature-engineering process. The feature vectors are trained and tested by classical machine-learning models like SVM.
In total, 54 features are generated by applying statistical functions (e.g., mean, variance and standard deviation) on three fields of each packet: \texttt{direction}, \texttt{packetsize}, \texttt{timeinterval}. 
To make the comparison fair, we added \texttt{dport} and \texttt{protocol} used by \system\ into the feature vector of \texttt{AppScanner} and applied one-hot encoding on them.
While embedding layer is used by \system\ to map the sparse \texttt{dport} values to a dense range, this technique cannot be used by the classical machine-learning models. Therefore, we apply Principal Component Analysis (PCA) on the feature vectors to achieve the similar effect, and reduce the dimension of the features to 100.
Next, we group the packets into packet window and aggregate the feature values across packets through vectorial sum to create feature vector. 
At the last step, the generated feature vector is normalized and fed into the SVM classifier.
Figure ~\ref{fig:svm_module} illustrates the workflow of our reproduced \texttt{AppScanner}. 

\subsubsection{Pure-traffic scenario} 
\label{sec:pure}

All of previous works on IoT traffic analysis~\cite{DBLP:journals/corr/ApthorpeRF17,DBLP:conf/infocom/SivanathanSGRWV17,miettinen2017iot, DBLP:conf/eurosp/TaylorSCM16,Mollers:2014:SPE:2627393.2627407} assume the flows of different devices can be differentiated. To assess how \system\ performs under this ``ideal'' network environment, we create this ``pure-traffic'' setting, that traffic belong to different devices can be clearly separated using the original device identifier, and compare \system\ and \texttt{AppScanner}. The feature vectors we create for the setting of Ethernet adversary, $FS_N$ and $FS_{NV}$ (see Table~\ref{tab:fs}), are evaluated separately for each model (CNN, LSTM, CNN-LSTM and SVM~\footnote{When $FS_{NV}$ is applied on SVM, \texttt{dport} and \texttt{protocol} are dropped.}). 


We set the packet window size $s$ to 100 and the moving step $m$ to 30. For each device, we extract 4,000 packet windows from 21 IoT devices in \texttt{Dataset-Pub}.
However, several devices provide fewer than 4,000 windows, so we obtained 41,405 traffic windows in total. 
We then split the traffic windows randomly into two datasets, with 80$\%$ packet windows for training and 20$\%$ for testing. A 10-fold cross-validation is performed to compute the metrics.

\begin{table}[t]
	\centering
	\caption{The accuracy of CNN, LSTM, CNN-LSTM and \texttt{AppScanner} (SVM) under the pure-traffic scenario. $FS_N$ and $FS_{NV}$ are feature vectors applied on each model.}
	\label{tab:metrics_pure_pure}
	\begin{tabular}{|l|l|l|l|l|l|l|}
	\hline
	\multirow{2}*{Model} & \multicolumn{3}{c|}{$FS_N$}  & \multicolumn{3}{c|}{$FS_{NV}$} \\
	\cline{2-7}
	~ & accuracy & precision & recall & accuracy & precision & recall \\
	\cline{1-7}
	CNN &  \textbf{97.1\%} & 97.4\% & \textbf{97.1}\% & 93.0\% & 93.2\% & 92.9\% \\
	\hline
	LSTM & 96.3\% & 96.6\% & 96.3\% & 95.5\% & 95.6\% & 95.5\% \\
	\hline
	CNN-LSTM & \textbf{97.1}\% & \textbf{98.1}\% & \textbf{97.1}\% & 95.0\% &\textbf{95.8\%} & 95.0\%\\
	\hline
	SVM  & 96.9\% & 96.9\% & 96.9\% & \textbf{95.7}\% & \textbf{95.8}\% & \textbf{95.8}\%\\
	\hline
	\end{tabular}
\end{table}


Table~\ref{tab:metrics_pure_pure} shows the performance of the four classification models. As we can see, all models performs well in identifying IoT devices using $FS_N$, reaching more than 96$\%$ in all three metrics. Among them, CNN-LSTM performs best. When $FS_NV$ is used, all the four models have a slight drop in their performances. The fact can be ascribed to missing features $dport$ and $protocol$. Among them, SVM performs best, with the three metrics exceeding 95$\%$, but the difference is quite small.


\subsubsection{Mixed-traffic scenario} 



\ignore{
\begin{figure}[t]
	\centering
	\includegraphics[width=0.85\columnwidth]{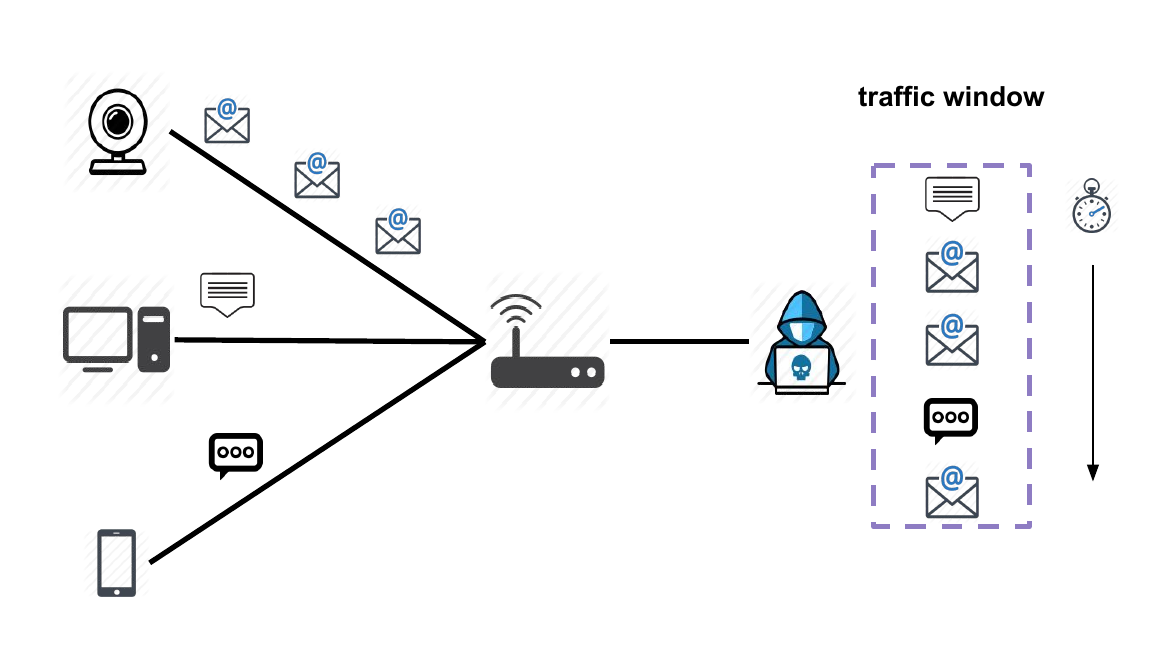}           
	\caption{Multi-device environment}
	\label{fig:multi_device}
	\vspace{-0.1in}
\end{figure}
}


In this scenario, we assume the NAPT or VPN is enabled by the gateway, so the traffic belong to the different devices could be mixed in the single packet window. 
Here we consider a classification result of a packet window is correct if it matches the device with the most packets (called \textit{dominant device}) in the window. 
For example, for a packet window with 80 packets from device $A$ and 20 packets from other devices, the right classification result should be $A$. Generally speaking, the dominant device should have the biggest impact on feature values.

Our DNN models are suitable for this task since it is able to produce a probability for each device through Softmax layer and the device with the highest probability is selected as output. 
Alternatively, we can select the devices whose probabilities exceed a threshold and output multiple devices. We plan to investigate this option in the future. 



\begin{figure}[t]
	\centering
	\includegraphics[width=0.6\columnwidth]{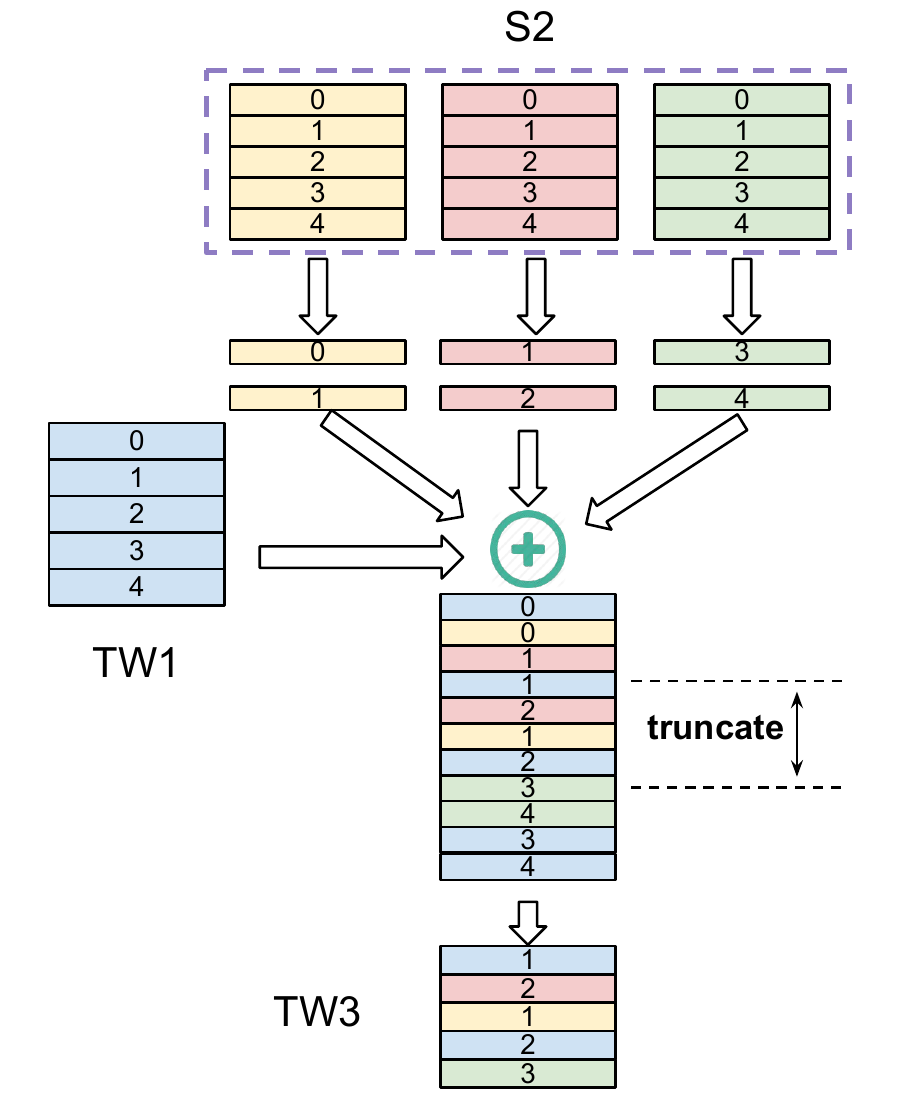}           
	\caption{The process of traffic mix.}
	\label{fig:mixed_gen}
    \vspace{-0.2in}
\end{figure}

\vspace{2pt} \noindent
\textbf{Data synthesis.} 
\texttt{Dataset-Pub} cannot be directly used for this evaluation task because the packets are well separated by device. We take a three-step procedure to derive the mixed-traffic dataset from \texttt{Dataset-Pub}. While generating mixed traffic from the real environment is another option, it is much more time-consuming and we discuss this issue in Section~\ref{sec:discussion}. 
The steps are listed below and illustrated in Figure ~\ref{fig:mixed_gen}. 

\begin{enumerate}
	\item We parse the traffic of all 21 IoT devices in \texttt{Dataset-Pub} into packet windows. Each window is named $TW_1$ and contains only one device's traffic. The set of $TW_1$ is named $s_1$.
	\item We mix the packets in each $TW_1$ with packets from other devices' windows. As a preparation step, we select $m$ windows from $m$ devices. The selected window is named $TW_2$ and the set of $TW_2$ is named $s_2$.
	\item Within each $TW_2$, a sequence of $n$ packets whose indices are in $[r, n+r]$ are replicated ($r$ is a random number) and assembled into a new window named $NW$. 
	\item $TW_1$ and the $m$ $NW$ are merged following the order of packet timestamp. 
	\item The new window is truncated to keep the same window size (assuming $w$), which consists of $w$ consecutive packets whose indices are in $[r, w+r]$ ($r$ is a random number). We name the new window $TW_3$. 
\end{enumerate}

After that, $TW_3$ contains the packets from a dominant device and $m$ other devices. The ratio between these two categories of packets (named \textit{mix ratio})  should have impact on the ultimate performance of \system. We experimented with two ratio values, 0.65 and 0.8. More concretely, assuming $TW_3$ has 100 packets, when mix ratio is set to 0.8, 80 packets should come from the dominant device. 

Same as the prior experiment, we selected 4,000 windows at maximum for each IoT device before traffic mixing and obtained 41,405 packet windows. We experimented with two training strategies to evaluate \system.
\begin{enumerate}
	\item \textit{Train ``mix'' and test ``mix''}. The training dataset has the mix ratio 0.8. The trained model is applied to two testing datasets of mix ratio 0.8 and 0.65.
	\item \textit{Train ``pure'' and test ``mix''}. 
    No traffic mix is performed on the training dataset. The  testing datasets have two mix ratio 0.8 and 0.65.
\end{enumerate} 




\begin{table}[t]
	\centering
	\caption{The performance of CNN, LSTM, CNN-LSTM and \texttt{AppScanner} (SVM) when training \textbf{``mix''} (mix ratio:\textbf{0.8}) and testing \textbf{``mix''} (mix ratio:\textbf{0.8}).}
	\label{tab:metrics_mix0.8_mix0.8}
	\begin{tabular}{|l|l|l|l|l|l|l|}
		\hline
		\multirow{2}*{Model} & \multicolumn{3}{c|}{$FS_N$}  & \multicolumn{3}{c|}{$FS_{NV}$} \\
		\cline{2-7}
		~ & accuracy & precision & recall & accuracy & precision & recall \\
		\cline{1-7}
		CNN &  95.6\% & 96.4\% & 95.6\% & 88.5\% & 89.5\% & 88.5\% \\
		\hline
		LSTM & \textbf{96.6\%} & 96.8\% & \textbf{96.6\%} & 93.9\% & 93.9\% & 93.9\%\\
		\hline
		CNN-LSTM & 96.2\% & \textbf{97.2\%} & 96.2\% & \textbf{94.2\%} & \textbf{94.3\%} & \textbf{94.2\%} \\
		\hline
		SVM  & 91.0\% & 91.2\% & 91.0\% &86.7\% & 86.6\% & 86.7\%\\
		\hline
	\end{tabular}
\end{table}

\begin{table}[t]
	\centering
	\caption{The performance of CNN, LSTM, CNN-LSTM and \texttt{AppScanner} (SVM) when training \textbf{``mix''} (mix ratio:\textbf{0.8}) and testing \textbf{``mix''} (mix ratio:\textbf{0.65}).}
	\label{tab:metrics_mix0.8_mix0.65}
	\begin{tabular}{|l|l|l|l|l|l|l|}
		\hline
		\multirow{2}*{Model} & \multicolumn{3}{c|}{$FS_N$}  & \multicolumn{3}{c|}{$FS_{NV}$} \\
		\cline{2-7}
		~ & accuracy & precision & recall & accuracy & precision & recall \\
		\cline{1-7}
		CNN &  95.0\% & 95.0\% & 95.0\% & 84.0\% & 85.8\% & 84.0\% \\
		\hline
		LSTM & \textbf{96.0\%} & \textbf{96.3\%} & \textbf{96.0\%} & \textbf{93.9\%}  & \textbf{93.9\%}  & \textbf{93.9\%} \\
		\hline
		CNN-LSTM & 95.5\% & 95.6\% & 95.5\% & 91.8\% & 92.1\% & 91.8\% \\
		\hline
		SVM  &91.5\%  & 91.9\% & 91.5\% &80.2\%  & 80.1\% & 80.2\%\\
		\hline
	\end{tabular}
\end{table}
\begin{table}[t]
	\centering
	\caption{The performance of CNN, LSTM, CNN-LSTM and \texttt{AppScanner} (SVM) when training \textbf{``pure''}  and testing \textbf{``mix''} (mix ratio:\textbf{0.8}).}
	\label{tab:metrics_pure_mix0.8}
	\begin{tabular}{|l|l|l|l|l|l|l|}
		\hline
		\multirow{2}*{Model} & \multicolumn{3}{c|}{$FS_N$}  & \multicolumn{3}{c|}{$FS_{NV}$} \\
		\cline{2-7}
		~ & accuracy & precision & recall & accuracy & precision & recall \\
		\cline{1-7}
		CNN &  \textbf{93.1\%} & \textbf{93.5\%} & \textbf{93.1\%} & 76.5\% & 80.5\% & 76.5\% \\
		\hline
		LSTM & 80.7\% & 82.9\% & 80.7\% & \textbf{90.3\%} & \textbf{90.6\%} & \textbf{90.3\%}  \\
		\hline
		CNN-LSTM & 91.5\% & 93.0\% & 91.5\% & 83.2\% & 85.1\% & 90.6\% \\
		\hline
		SVM  & 67.9\% & 70.9\% & 67.9\% & 67.8\% & 71.2\% & 67.8\%\\
		\hline
	\end{tabular}
\end{table}
\begin{figure}[t]
	\centering
	\includegraphics[width=0.75\columnwidth]{pic/confusion_matrix.png}           
	\caption{The confusion matrix of CNN on device identification. The number beside each axis is the index of IoT device.}
	\label{fig:confusion_matrix}
	\vspace{-0.2in}
\end{figure}
\begin{table}[t]
	\centering
	\caption{The performance of CNN, LSTM, CNN-LSTM and \texttt{AppScanner} (SVM) when training \textbf{``pure''}  and testing \textbf{``mix''} (mix ratio:\textbf{0.65}).}
	\label{tab:metrics_pure_mix0.65}
	\begin{tabular}{|l|l|l|l|l|l|l|}
		\hline
		\multirow{2}*{Model} & \multicolumn{3}{c|}{$FS_N$}  & \multicolumn{3}{c|}{$FS_{NV}$} \\
		\cline{2-7}
		~ & accuracy & precision & recall & accuracy & precision & recall \\
		\cline{1-7}
		CNN & \textbf{86.6\%} & \textbf{88.2\%} & \textbf{86.6\%} & 68.4\%  & 73.0\%  & 68.4\% \\
		\hline
		LSTM & 71.5\% & 76.5\% & 71.5\% & \textbf{86.0}\% & \textbf{87.1}\% & \textbf{86.0}\% \\
		\hline
		CNN-LSTM &  83.5\% & 87.3\% & 83.5\% & 72.5\% & 77.1\% & 72.5\% \\
		\hline
		SVM  & 58.0\%  &62.6\% & 58.0\%  & 58.2\% & 63.1\% & 58.2\% \\
		\hline
	\end{tabular}
\end{table}

%
%

\vspace{2pt} \noindent
\textbf{Evaluation results.} 
We first evaluate how \system\ performs in an NAPT environment. The feature vector is $FS_N$. 

When the model trained with the mixed traffic, all four models achieve high accuracy, precision and recall, as shown in Table ~\ref{tab:metrics_mix0.8_mix0.8}. When mix rate of testing dataset is set to 0.8, the accuracies for CNN, LSTM and CNN-LSTM are 95.6\%, 96.6\% and 96.2\%, outperforming SVM (91.0\%). The gap is similar when we reduce the mix rate of testing dataset to 0.65 (Table~\ref{tab:metrics_mix0.8_mix0.65}).

\begin{table}[]
		\caption{The performance of CNN for device identification individually, when training ``\textbf{pure}'' and testing ``\textbf{mix}'' (mix ratio: \textbf{0.8}). \#S: number of samples.}
	\label{tab:LSTM_dev_identify}
\begin{tabular}{|l|l|l|l|l|}
	\hline
	\textbf{Device}                                                          & \textbf{Accuracy} & \textbf{Precision} & \textbf{Recall} & \textbf{\#S} \\ \hline
	iHome                                                                    & 93.0\%            & 97.3\%             & 93.0\%          & 158          \\ \hline
	Samsung SmartCam                                                         & 100.0\%            & 89.3\%             & 100.0\%          & 800          \\ \hline
	Dropcam                                                                  & 99.6\%            & 99.9\%             & 99.6\%          & 800          \\ \hline
	Belkin Wemo Motion Sensor                                                & 97.1\%            & 56.4\%             & 97.1\%          & 327          \\ \hline
	\begin{tabular}[c]{@{}l@{}}Withings Smart Scale\end{tabular}           & 0.0\%            & 0.0\%             & 0.0\%          & 5            \\ \hline
	Triby Speaker                                                            & 98.7\%            & 91.0\%             & 98.7\%          & 254          \\ \hline
	Insteon Camera                                                           & 99.5\%            & 99.9\%             & 99.5\%          & 800          \\ \hline
	Nest Protect Smoke Alarm                                                 & 100.0\%           & 100.0\%             & 100.0\%         & 9            \\ \hline
	Nest Dropcam                                                             & 98.4\%            & 98.4\%             & 100.0\%          & 67           \\ \hline
	Light Bulbs LiFX  Smart Bulb                                             & 100.0\%            & 100.0\%            & 100.0\%          & 448          \\ \hline
	Amazon Echo                                                              & 99.8\%            & 99.9\%             & 99.8\%          & 800          \\ \hline
	Withings Smart Baby Monitor                                              & 82.5\%            & 100.0\%            & 82.5\%          & 94           \\ \hline
	Netatmo Welcome                                                          & 100.0\%            & 98.2\%             & 100.0\%          & 800          \\ \hline
	\begin{tabular}[c]{@{}l@{}}TP-Link Day Night Cloud Camera\end{tabular} & 98.8\%            & 99.4\%             & 98.8\%          & 800          \\ \hline
	PIX-STAR Photo Frame                                                     & 99.0\%            & 100.0\%             & 99.0\%          & 113          \\ \hline
	TP-Link Smart Plug                                                       & 98.8\%            & 92.1\%             & 98.8\%          & 100          \\ \hline
	Withings Aura Smart Sleep Sensor                                         & 89.8\%            & 100.0\%             & 89.8\%          & 800          \\ \hline
	Belkin Wemo Switch                                                       & 4.8\%            & 65.0\%             & 4.8\%          & 282          \\ \hline
	Netatmo Weather Station                                                  & 100.0\%           & 100.0\%             & 100.0\%         & 144          \\ \hline
	Blipcare Blood Pressure Meter                                            & 0.0\%             & 0.0\%              & 0.0\%           & 0            \\ \hline
\end{tabular}
\end{table}

Next, we evaluate how \system\ performs under NAPT+VPN environment, in which the feature vector $FS_{NV}$.
Following the same experiment steps, we train models with the mixed traffic. When the mix ratio for testing dataset is 0.8, CNN-LSTM performs best and leading SVM with around 8\% in  all three metrics (Table ~\ref{tab:metrics_mix0.8_mix0.8}). 
When the mix ratio of testing dataset is reduced to 0.65, the gap between SVM and DNN models is widened. LSTM is able to outperform SVM by more than 13\% in all three metrics (Table ~\ref{tab:metrics_mix0.8_mix0.65}). While it is expected the best performance should be achieved when the mix ratios are the same for training and testing dataset, our result suggests DNNs are more resilient to the such variance.

From the perspective of an adversary, knowing the mix ratio ahead (or even roughly) is quite difficult, so we expect that the strategy of training with ``pure'' data is more likely to be adopted by the adversary. 
Table ~\ref{tab:metrics_pure_mix0.8} and Table ~\ref{tab:metrics_pure_mix0.65} show the evaluation result when the training dataset is ``pure''. 
From the table, we can see CNN continues to perform well, with 93.1\% accuracy when testing mix ratio is 0.8 and 86.6\% when the ratio is 0.65. A larger performance decrease happens on two other DNNs but the decrease for SVM is much more significant: the result drops with the range of 15\% to 30\%.

Table~\ref{tab:LSTM_dev_identify} lists the evaluation result for individual device under CNN model. We can see the model works well on most of the involved IoT devices except for Withings Smart Scale, Belkin Wemo Motion Sensor (low precision) and Belkin Wemo Switch (the number of packets is too small for Blipcare Blood Pressure Meter to fill a packet so all three metrics are 0). The bad performance on Withings Smart Scale can be ascribed to the shortage of training and testing samples in the \texttt{Dataset-Pub}. From the perspective of an Ethernet attacker, the traffic generated by Belkin Wemo Motion Sensor and Belkin Wemo Switch are so similar that it is hard to distinguish between them. Figure~\ref{fig:confusion_matrix} shows the confusion matrix for device identification under the same setting. 




In summary, our evaluation results show DNNs like CNN, LSTM and CNN-LSTM in deed achieve better effectiveness, especially under the more practical scenarios, like when NAPT or VPN is enabled and the adversary has no a priori knowledge about the traffic distribution of the targeted smart home. Feature engineering is much more difficult under those complex environment and DNN's capability of automatic feature representation learning can address this challenge. 
Comparing to CNN, the performance degrades more significantly for LSTM and CNN-LSTM when training with ``pure'' traffic. This finding could be explained by the nature of our data. In the data synergy stage, we randomly merge packets from different devices, which could break the latent temporal relation between packets. 
As such, the difference between training and testing dataset could appear more prominent to RNN-based models.




\ignore{
\vspace{2pt} \noindent
\textbf{Impact of packet window size ($s$).} 
During the prior experiments, we set a fixed window size ($s$) at 100 packets. Here we adjust $s$ from 100 to 200 to evaluate its impact on performance.
After increasing $s$, the three DNN models achieve higher accuracies when the training dataset consists of pure traffic in most cases (except LSTM-RNN for NAPT+VPN). The DNN also gains better accuracy when for the mixed training traffic (except LSTM-RNN for NAPT). On the other hand, SVM's performance is worsen when more packets are included, suggesting the noise introduced by the new packets cannot be overlooked. 

\zl{do we have another window size, say, 50?}

%
\begin{figure}
	\subfigure[window size = 100]{
		\centering
		\includegraphics[width=0.75\columnwidth]{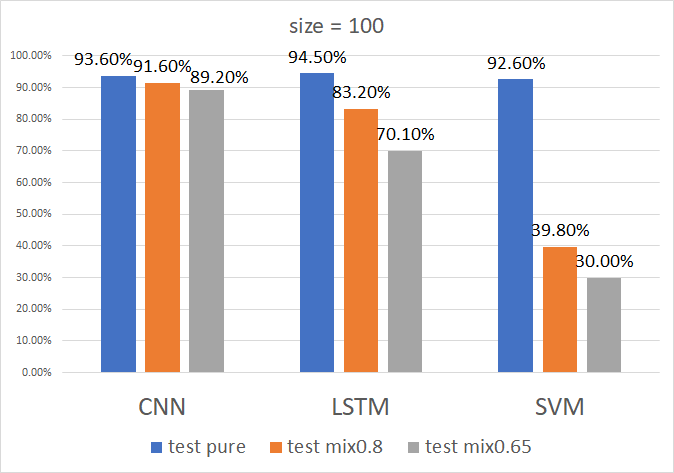}
		\label{fig:influence_napt:a}

}
\subfigure[window size = 200]{
		\centering
		\includegraphics[width=0.75\columnwidth]{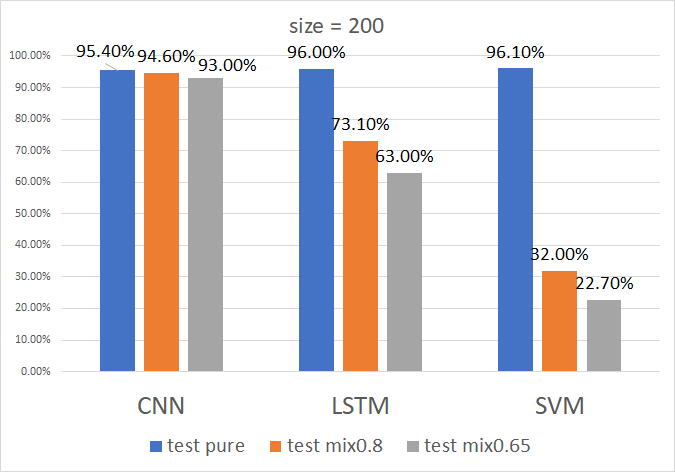}
		\label{fig:influence_napt:b}
}
\caption{Performance of different window size in NAPT environment.\zl{taking too much space}}
\label{fig:windowsize_napt}
\vspace{-0.2in}
\end{figure}

\begin{figure}
	\subfigure[window size = 100]{
			\centering
			\includegraphics[width=0.75\columnwidth]{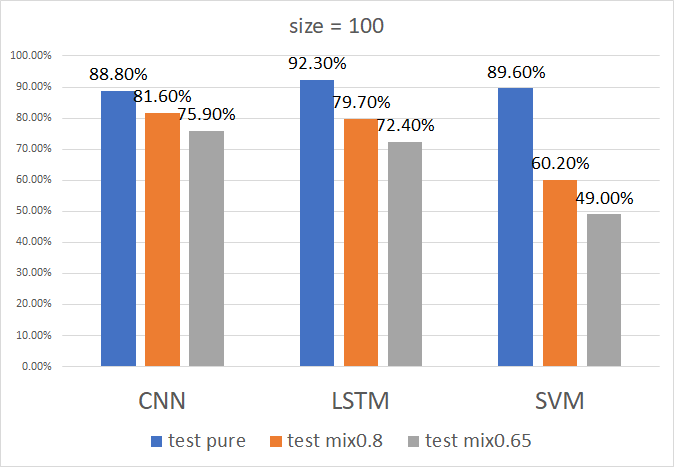}
			\label{fig:influence_vpn:a}
	}

	\subfigure[window size = 200]{

			\centering
			\includegraphics[width=0.75\columnwidth]{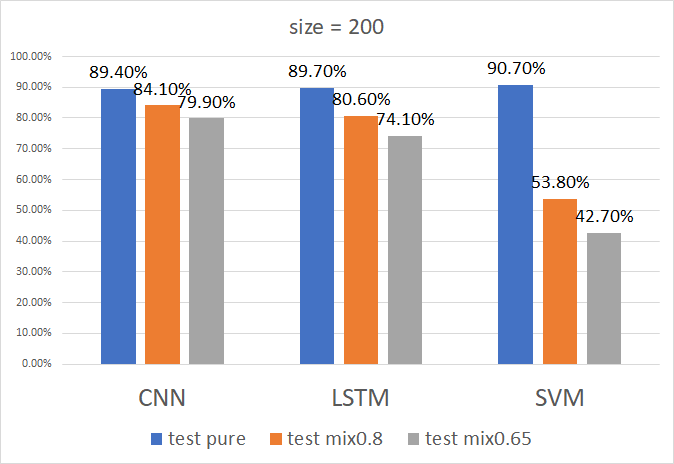}
			\label{fig:influence_vpn:b}

	}
	\caption{Performance of different window size in NAPT+VPN environment.}
	\label{fig:windowsize_vpn}
	\vspace{-0.2in}
\end{figure}
}

\subsubsection{Wireless sniffing} 
We use another dataset \texttt{Dataset-WL} to assess the effectiveness of \system\ when only the Wi-Fi traffic is available to the adversary. Recall that the original device identifier is preserved under this setting, so we did not test the setting of traffic mix. On the other hand, due to the encryption of WPA or WPA2, less features can be used (see $FS_{WL}$ in Table~\ref{tab:fs}). 

\begin{table}[t]
	\centering
	\caption{The overall performance of CNN, LSTM, CNN-LSTM in wireless sniffing.}
	\label{tab:metrics_wireless_sniffing}
	\begin{tabular}{|l|l|l|l|}
		\hline
		\textbf{Model} & \textbf{Acuracy} & \textbf{Precision} &\textbf{Recall} \\
		\hline
		CNN & 94.9\% & 94.6\% & 94.9\%\\
		\hline
		CNN-LSTM & 96.2\% & 96.0\% & 96.2\%\\
		\hline
		LSTM & 98.9\% & 97.7\% & 98.9\%\\
		\hline
	\end{tabular}
\end{table}

Though the features we can use are few, all three models (CNN, LSTM, CNN-LSTM) achieve quite high accuracy. Among them, LSTM performs best, with an accuracy score over 98\%.  Table~\ref{tab:metrics_wireless_sniffing} summarizes the result for the three DNN models. In Table~\ref{tab:lstm_wireless} shows the evaluation results for each device. As we can see, every IoT device can be recognized with quite high accuracy (the lowest is 96.4\% for Amazon Echo Dot). Interestingly, the three non-IoT devices are also recognizable with good accuracy.
\begin{table}[t]
	\centering
	\caption{LSTM performance in wireless sniffing attack.}
	\label{tab:lstm_wireless}
	\begin{tabular}{|l|l|l|l|l|}
		\hline
		\textbf{Device} & \textbf{Accuracy } & \textbf{Precision} & \textbf{Recall} & \textbf{\#S} \\
		\hline
		Google Home & 100.0\% & 100.0\% & 100.0\% & 95 \\
		\hline
		Amazon Echo Dot & 96.4\% & 100.0\% & 96.4\% & 111 \\
		\hline
		Tmall Assistant & 99.7\% & 98.6\% & 99.7\% & 384 \\
		\hline
		Xiaobai camera & 100.0\% & 100.0\% & 100.0\% & 12\\
		\hline
		Kongke Switch & 100.0\% & 100.0\% & 100.0\%  & 93 \\
		\hline
		Sony Phone	& 99.2\% & 89.2\% & 99.2\% & 142 \\
		\hline
		Motorola Phone & 94.3\% & 89.3\% & 94.3\% & 21 \\
		\hline
		Netcore Router &  98.7\% & 98.1\% & 98.7\% & 800 \\
		\hline
	\end{tabular}
\end{table}

\subsection{Activity Inference}


In this subsection, we report our evaluation results of activity inference, using the dataset \texttt{Dataset-Act} collected by ourselves. As described in Section~\ref{sec:activity}, we tested 8 IoT devices belong to 3 categories and simulated 20 user behaviors. 

Due to the space limit, here we focus on the experiments under NAPT environment. The feature set we used is $FS_N$. 
We totally collected more than 1,058,000 packets to cover all the activities. After data pre-processing, 315,080 packets are preserved for training and testing. 
Because a user activity usually results in a traffic burst, in which the interval between request and response is small, 
we reduce the packet window size $s$ and the moving step $m$ to 50 and 25. In the end, we obtained 11,181 packet windows. Same as the experiment of device identification, we assign 80\% packet windows to training dataset and the remaining 20\% to testing dataset. For this experiment, we only test the ``pure-traffic'' scenario so the traffic are separated based on devices, because the packet windows of this dataset are much fewer than \texttt{Dataset-Pub} (41,405 windows).

\begin{table}[t]
	\centering
	\caption{The performance of CNN, LSTM, CNN-LSTM on Activity Inference.}
	\label{tab:metrics_activity}
	
	\begin{tabular}{|l|l|l|l|}
		\hline
		\textbf{Model} & \textbf{Accuracy} & \textbf{Precision} & \textbf{Recall} \\
		\hline
		CNN & 87.9\% & 86.1\% & 87.9\%\\
		\hline
		LSTM & \textbf{92.8\%} & \textbf{92.9\%} & \textbf{92.8\%}\\
		\hline
		CNN-LSTM & 91.7\% & 91.8\% & 91.7\% \\
		\hline
	\end{tabular}
\end{table}

\begin{table}[t]
	\centering
	\caption{Activity inference result of LSTM for individual device.}
	\label{tab:activity_infer_performance}
	\scalebox{0.92}[0.92]{%
		\begin{tabular}{|l|l|l|l|l|l|}
			\hline
			\textbf{Device} &\textbf{Activity} & \textbf{Accuracy} & \textbf{Precision} & \textbf{Recall} & \textbf{\#S}                                                                           \\
			\hline
			\multirow{4}*{Amazon Echo Dot} &			
			Standby & 75.0\% & 60.0\% & 75.0\% & 4
			 \\ \cline{2-6}
			~& Control Philips bulb &10.0\% & 25.0\% & 10.0\% & 46\\
			\cline{2-6}
			~& Control Wemo switch & 71.0\% & 52.0\% & 71.0\% & 56 \\
			\cline{2-6}
			~& Control TP-Link switch &59.0\% & 7\% & 59.0\% & 22\\
			\cline{2-6}
			\hline
			\multirow{4}*{Tmall Assistant} & Standby &97.0\% & 100.0\% & 97.0\% & 23 \\
			\cline{2-6}
			~ & Wakeup & 97.0\% & 88.0\% & 97.0\% & 400\\
			\cline{2-6}
			~ & Control Tospo bulb & 52.5\% & 81.6\% & 52.5\% &162\\
			\cline{2-6}
			~ & Play song & 82.6\% & 95.0\% & 82.6\% &59\\
			\hline
			\multirow{3}*{Google Home} & Standby & 99.4\% & 93.6\% & 99.4\% & 180\\
			\cline{2-6}
			~ &Wakeup & 96.8\% & 99.7\% & 96.8\% & 400\\
			\cline{2-6}
			~ &Ask question &91.8\% & 95.7\% & 91.8\% & 73\\
			\hline
			
			SmartThings& Control Philips Hue & 100.0\% & 92.0\% &100.0\% & 11 \\			
			\hline
			\multirow{3}*{Xiaomi Hub} & control Xiaomi ring & 100.0\% & 30\% & 100.0\% & 3\\
			\cline{2-6}
			~ & Live monitor(WLAN) & 99.4\% & 100.0\% & 99.4\% & 160\\
			\cline{2-6}
			~ &Live monitor(WAN) & 100.0\% & 100.0\% & 100.0\% & 818\\
			\hline
			Belkin Wemo Switch & turn on/off & 100.0\% & 100.0\% & 100.0\% & 25\\		
			\cline{2-6}
			\hline
			
			Tmall box & Play movies &99.4\% & 98.2\% & 99.4\% & 169\\
			\hline	
			\multirow{2}*{Orvibo Switch} &Standby & 100.0\% & 100.0\% & 100.0\% & 1\\
			\cline{2-6}
			~&Turn on/off & 100.0\% & 100.0\% & 100.0\% & 12\\
			\cline{2-6}
			
			\hline
		\end{tabular}
	}
\end{table}

Overall, our CNN, LSTM and CNN-LSTM models achieve accuracies of 87.9$\%$, 92.8$\%$ and 91.7$\%$ respectively, as shown in Table~\ref{tab:metrics_activity}.  Table~\ref{tab:activity_infer_performance} specifies the result for each activity under LSTM model.
Among them, 3 behaviors have relatively low accuracies, including controlling the Tospo bulb through Tmall Assistant
(52.5\%), controlling TP-link switch with Amazon Echo Dot (59.0\%) and controlling Philips bulb with Amazon Echo Dot (10.0\%). Interestingly, all the three cases are related to the combination of smart assistant and accessory. After diagnosing the root cause, we found they share similar patterns in network communications and device interactions, which can be summarized as below: 

\ignore{
\begin{enumerate}
	\item Tmall Assistant controls the bulb: 47.9\%
	\item Amazon Echo Dot controls Wemo switch: 48\%
	\item Amazon Echo Dot controls Philips Hue: 23\%
\end{enumerate}
}


\begin{enumerate}
	\item The user spells a wakeup voice command, like ``Alexa'' or ``Hello! Tmall Assistant''.
	\item After recognizing the command, the smart assistant transits to the working mode and listens to the next command.
	\item The user spells a controlling command like ``Turn off the TP-Link switch'' or ``Light up the bulb''.
	\item The smart assistant receives the command, identifies the target IoT device within the smart home network and sends the control packets to the device. (Echo Dot uses Wi-Fi and Tmall Assistant uses BLE.)
	\item The smart assistant communicates with the remote server and exchanges information like activity logs and device status.
\end{enumerate}

The steps (4) and (5) generate network traffic, but only communication in step (5) can be captured by the Ethernet adversary, which turns out to be quite similar each time. The communication in step (4) is more distinctive but the packets can only be captured by a wireless adversary. Unless the two adversaries can collude, accurate activity inference is quite difficult.



\ignore{
\begin{figure}
	\begin{minipage}[t]{0.5\columnwidth}
		\centering
		\includegraphics[width=1.6in]{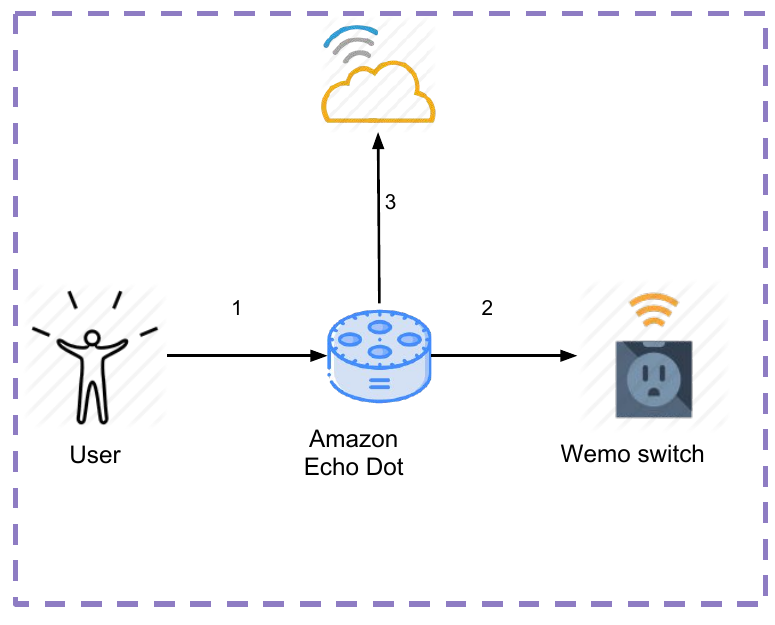}
		\caption{Amazon Echo \\ Dot and Belkin Wemo Hue}
		\label{fig:echo:a}
	\end{minipage}%
	\begin{minipage}[t]{0.5\columnwidth}
		\centering
		\includegraphics[width=1.6in]{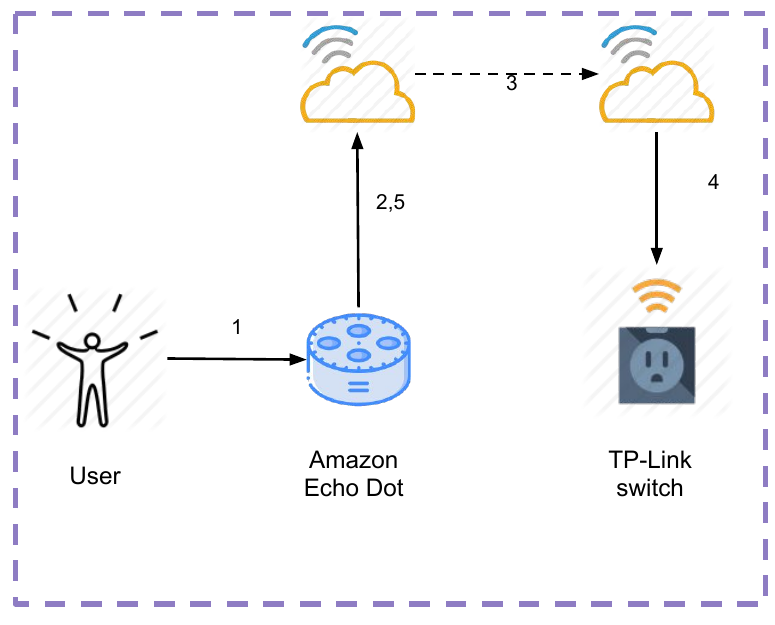}
		\caption{Amazon Echo \\ Dot and TP-Link switch}
		\label{fig:echo:b}
	\end{minipage}
\vspace{-0.2in}
\end{figure}
}

\ignore{
\subsubsection{Hybrid approach to activity inference.} 

As the limitation of an Ethernet attacker, we propose a hybrid way of combining various evidence together to enhance the effectiveness of an inference attack. The approach has the following key points.


\begin{itemize}
	\item \textit{Identify the device through recognizing the heartbeat packets.} As Section ~\ref{sec:observation} says, heartbeat packets are usually sent from a device to the remote server, so as to keep the connection between two parties. The response packets from the server are usually typical as well. The best way of collecting heartbeat packets is to monitor the network traces in a standby mode (i.e. waiting for the command and currently no workload)
	
	\item \textit{Combine wireless sniffing results.} For those chained behaviors, network traces collected outside the router may be not enough, like the above three examples. To help improve the precision of the inference, Ethernet sniffing and wireless sniffing can be combined.

\end{itemize}

We evaluate the hybrid approach with two devices involved, Amazon Echo Dot and Belkin Wemo switch. The experiment setting is as Figure ~\ref{fig:adversary} shows without other devices interference. We bind Belkin Wemo switch and Amazon Echo Dot together to turn on/off the switch for 10 times with a 10-second gap and collect all the traffic (both the wireless traffic and that passing through the router).



\textbf{Evaluation.}
After removing the control and management frames, we align the remaining data frames with packets collected by the Ethernet sniffing attacker by their timestamps and finally obtain a group of frames that are potentially used for Belkin wemo control. 
Among 10 control behaviors, 2 intact frame groups are pinpointed and verified by checking the MAC address. For the 8 failure cases, we found it is the packet loss of the sniffer that leads to the incomplete frame groups. Though only 2 of 10 repetitions successfully affirm the controlling behavior between Amazon Echo Dot and Belkin Wemo switch, we believe the performance can be improved with the help of a equipment with lower Packet Loss Ratio.
}
}
	\subsection{Case Studies}
\label{sec:case_study}

\vspace{2pt} \noindent
\textbf{Scarce traffic.}
Several IoT devices have a low traffic volume, such as \texttt{orvibo plug} and \texttt{tplink plug}. Therefore, the packets generated by them may be ``overwhelmed'' by other devices in the same traffic window.  
Take \texttt{orvibo plug} as an example. It has a three-TCP-packet sequence appearing multiple times. Their \texttt{frame lengths} are 224, 54, 240 bytes in two seconds correspondingly. 
But those numbers are not unique. According to our analysis, packets with the same sizes are generated by other devices so our classification models could be confused.
In particular, \texttt{360 camera} produces UDP packet of 224 bytes very frequently. \texttt{xiaomi hub} and \texttt{xiaomi tablet} generate TCP packets of 54 bytes consistently at high speed 
(15 packets in 0.1 second).
As such, the three consecutive packets of \texttt{orvibo plug} are likely to be separated by those packets from other devices, leading to wrong classification results, especially in VPN configuration when protocol information is missing. 
This could explain why LSTM-RNN models perform worse in identifying IoT devices of small traffic volume in VPN environment. 

\vspace{2pt} \noindent
\textbf{Effectiveness of bidirectional LSTM.}
Section~\ref{sec:evaluation} shows the accuracy of bidir LSTM model is better than the basic LSTM model in most cases. Below we use one example to explain this performance difference. 
Through our manual check of traffic from different devices, we find packets of the size 66 bytes are commonly sent to the server in a sequence by \texttt{Echo Dot} and \texttt{Google Home}. 
However, the responses from the server are different between two devices. For \texttt{Echo Dot}, most of the responses are of a size 1388 bytes while for \texttt{Google Home}, most of them are TLS packets with the size of 108 bytes and 105 bytes. 
Compared to a basic LSTM, a bidirectional LSTM can utilize the later packet (1388-byte or 105-byte) to help classify the previous packets (66-byte). As such, adding the information from ``future'' could improve the chance of correctly classifying packets.

	\section{Discussion and limitation}
\label{sec:discussion}

\noindent \textbf{Packet-level identification.} Different from previous work that identifies devices based on network flows or traffic windows, we assign the device label to every packet generated within a period. The motivation for doing packet-level classification is that the running status of IoT devices can be more accurately identified and the device information can be obtained more promptly.
For example, if packets are observed sparsely, the device might be in sleeping or standby mode. Otherwise, a dense packet sequence indicates the device is busy running a task. 
This is critical in some cases like camera monitoring, if a thief knows when a network camera is transmitting bulk data, he can infer whether the host is monitoring the house and decide the best time to sneak into the house. 
Additionally, we can combine the predicted labels of all packets in a time window to make better inference of IoT devices.

\noindent \textbf{VPN connections.} During the experiment, we only establish our VPN connection using UDP protocols. Another option -- TCP(TLS) is not tested. The main reason is that UDP is the default protocol adopted by the \texttt{openVPN} service and is the most-widely protocol used by VPN providers due to its low latency~\cite{url_udp_tcp} compared with TCP. 
VPN using TCP requires modification of our packet labeling algorithm in Section~\ref{sec:traffic_preprocess} and we decide to leave it as our future work.

\noindent \textbf{Behavior identification.}We focused on device identification during evaluation, while previous works also explored user behavior identification~\cite{Acar2018PeekaBooIS}. We did not experiment with behavior identification because the labeling cost is high given that we have large datasets with millions of IoT packets. On the other hand, we believe our models can be applied to this scenario if we have enough training samples. We will explore approaches that can generate labeled behavioral datasets efficiently.

	\section{Conclusion}
\label{sec:conclusion}
In this paper, we systematically evaluated the effectiveness of traffic analysis in a smart home environment, even when traffic fusion like NAPT and VPN are enabled and non-IoT and IoT devices are both active. 
By exploiting the dependency between packets through DNN models like LSTM-RNN, we show it is possible to achieve high accuracy in device identification, even under the complex network environment as described above. 

Our result suggests the network communications of IoT devices do have serious privacy implications, even under encryption and traffic fusion. We believe more research should be done to better understand the privacy issues in smart home network and mitigate such issues. To facilitate the research in this domain, we will release the data and our models.
	\bibliographystyle{ACM-Reference-Format}
	\bibliography{ref}
	\begin{appendices}
\section{Traffic Patterns of IoT Devices}
\label{sec:traffic_pattern}

\begin{figure*}[p]
	\begin{minipage}[t]{1.0\columnwidth}
		\centering
		\includegraphics[width=2.5in,height=1.0in]{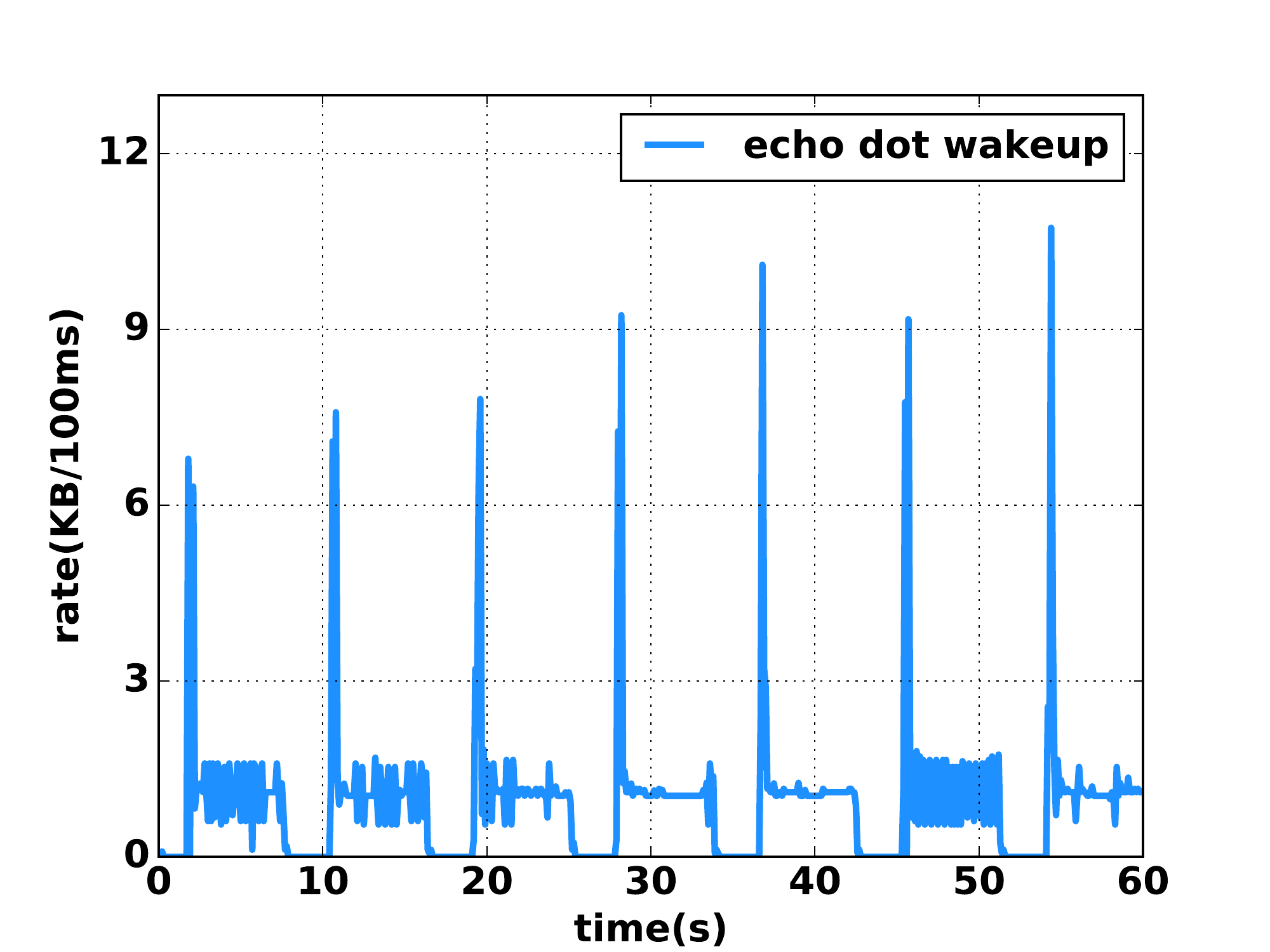}
		\caption{Echo dot.}
		\label{fig:echodot}
		\vspace{-0.1in}
	\end{minipage}%
	\begin{minipage}[t]{1.0\columnwidth}
		\centering
		\includegraphics[width=2.5in,height=1.0in]{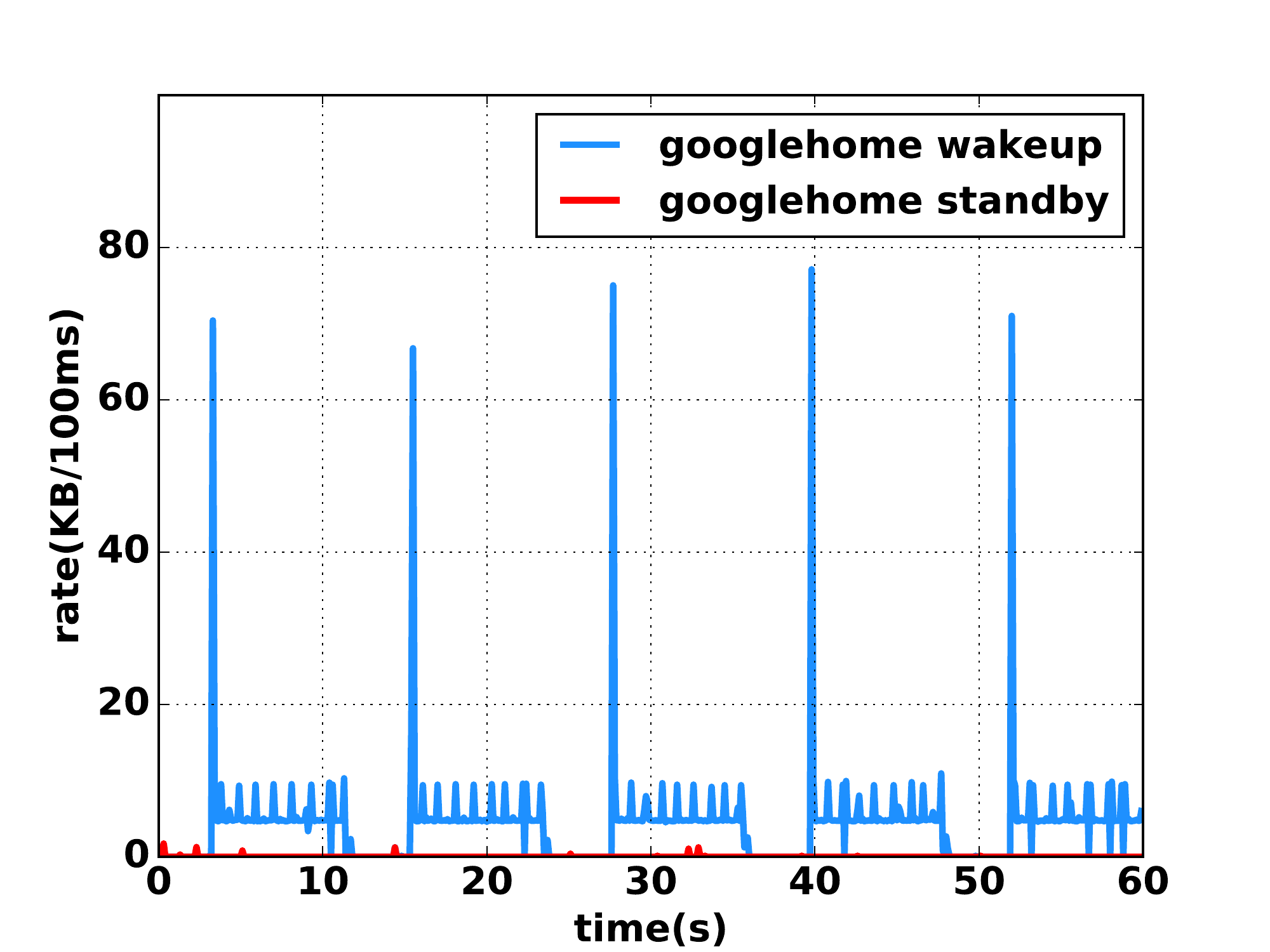}
		\caption{Google Voice Assistant}
		\label{fig:google_voice}
		\vspace{-0.1in}
	\end{minipage}%
\end{figure*}

Figure~\ref{fig:echodot} and figure~\ref{fig:google_voice} show the traffic patterns when waking Echo Dot and Google Home up.


\section{Performance in Noisy Environment}
\label{sec:appendix_performance_pureIoT}

\begin{figure*}[p]
	\begin{minipage}[t]{1.0\columnwidth}
    	\centering
    	\includegraphics[width=0.85\columnwidth]{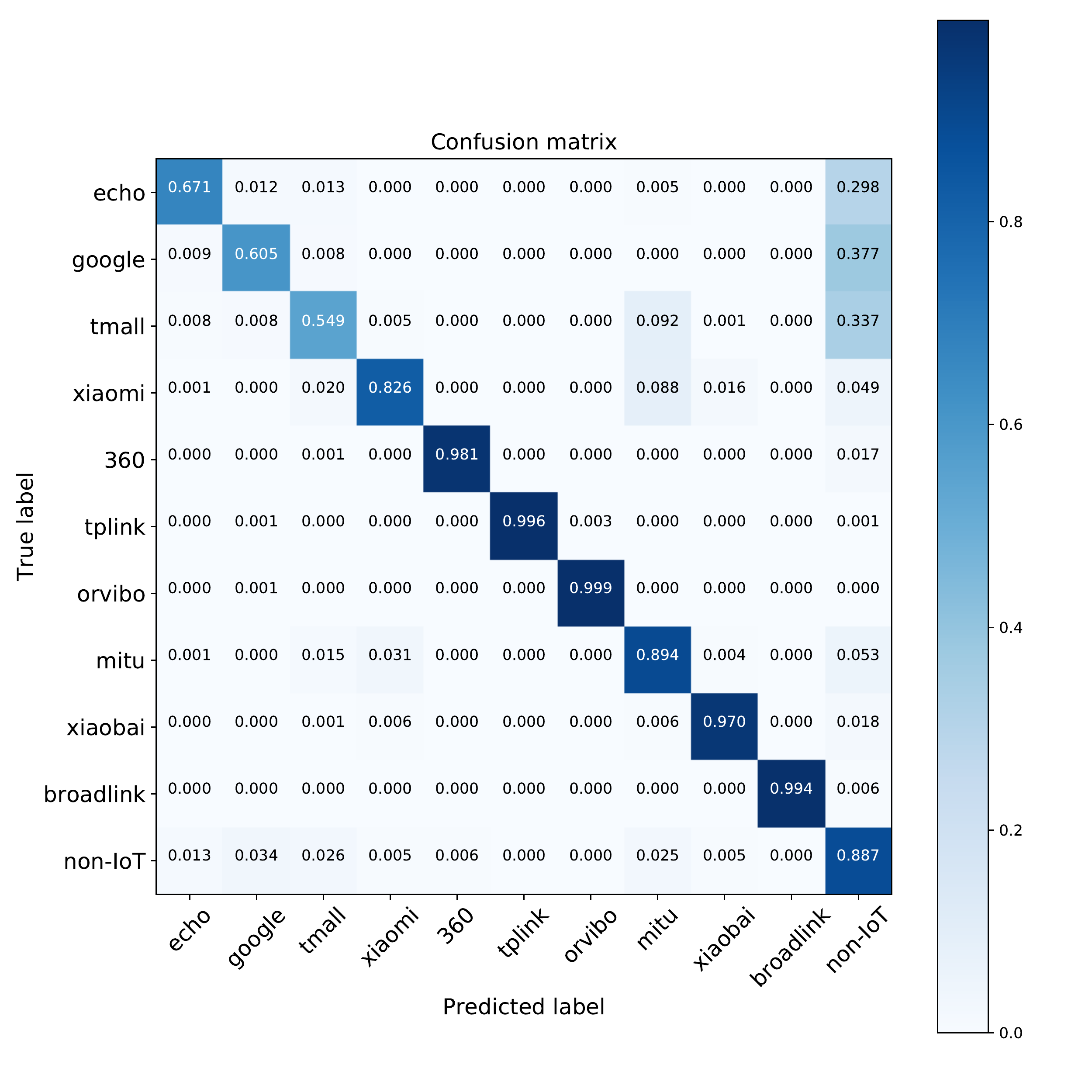} 
    	\caption{Confusion matrix of RF (noisy+NAPT).}
    	\label{fig:cm_rf_noisy_napt}
    \end{minipage}
    \begin{minipage}[t]{1.0\columnwidth}
    	\centering
    	\includegraphics[width=0.85\columnwidth]{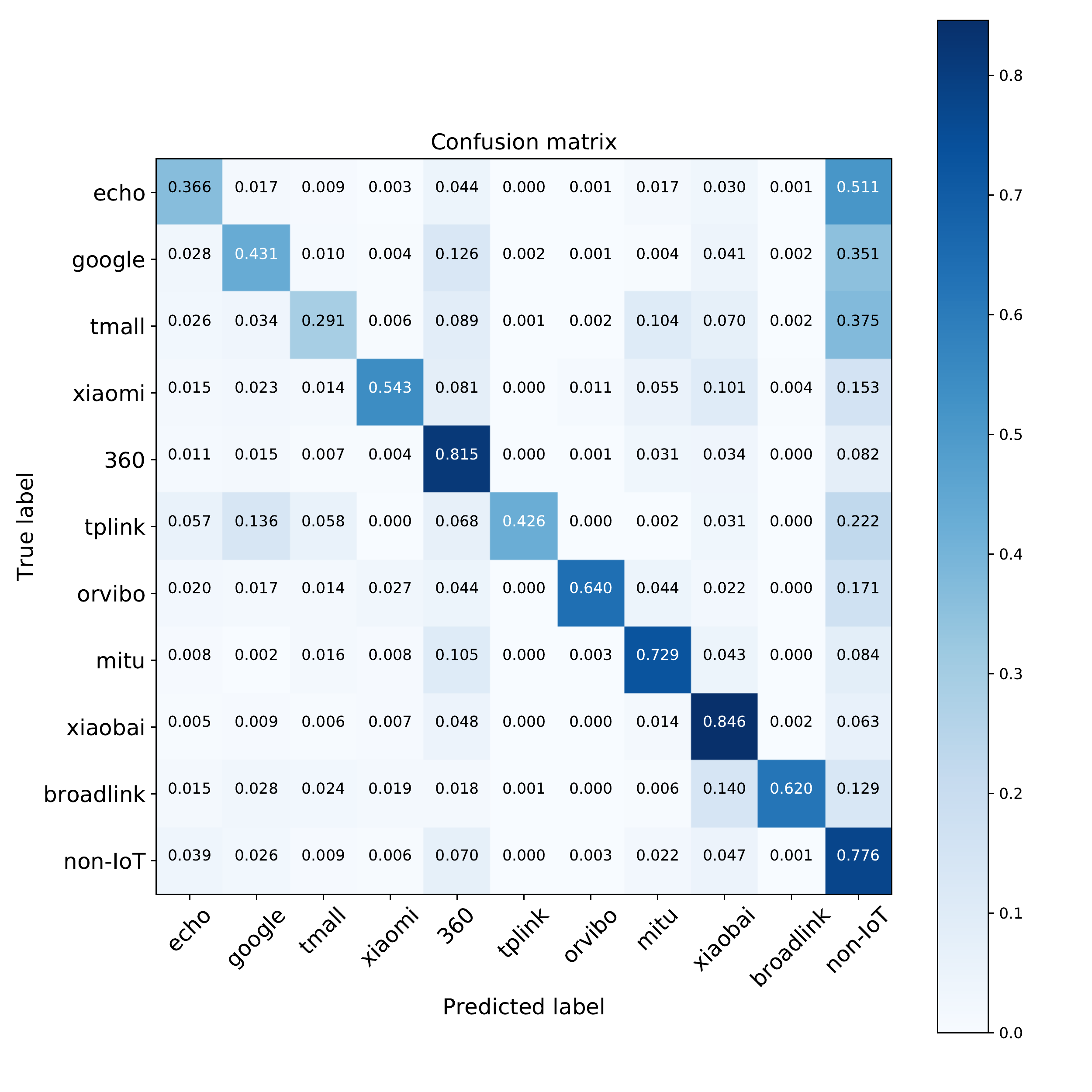} 
    	\caption{Confusion matrix of RF (noisy+VPN).}
    	\label{fig:cm_rf_noisy_vpn}
    \end{minipage}
\end{figure*}

\begin{figure*}[p]
	\begin{minipage}[t]{1.0\columnwidth}
    	\centering
    	\includegraphics[width=0.85\columnwidth]{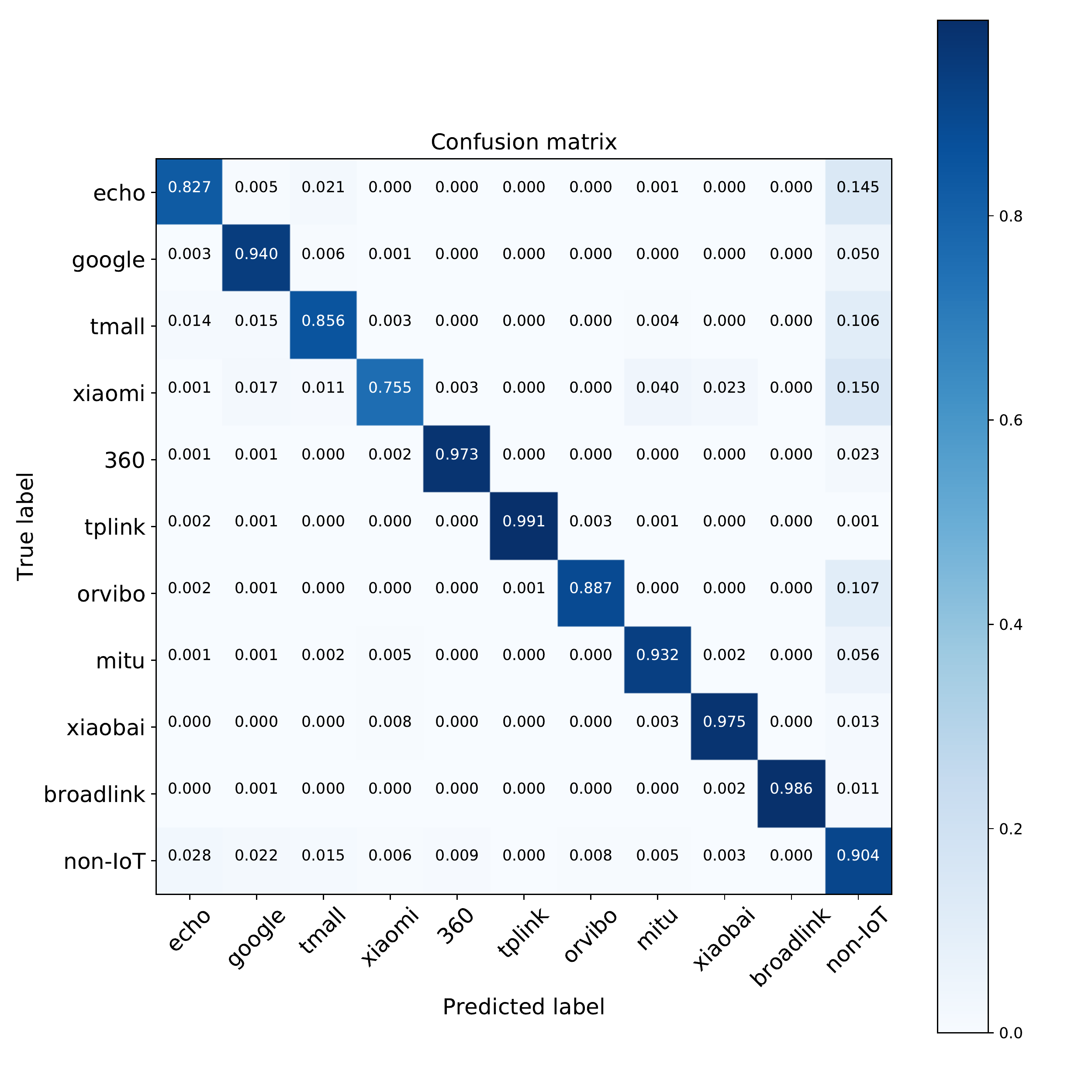} 
    	\caption{Confusion matrix of BLSTM (noisy+NAPT).}
    	\label{fig:cm_blstm_noisy_napt}
    	    	\vspace{-0.2in}

    \end{minipage}
    \begin{minipage}[t]{1.0\columnwidth}
    	\centering
    	\includegraphics[width=0.85\columnwidth]{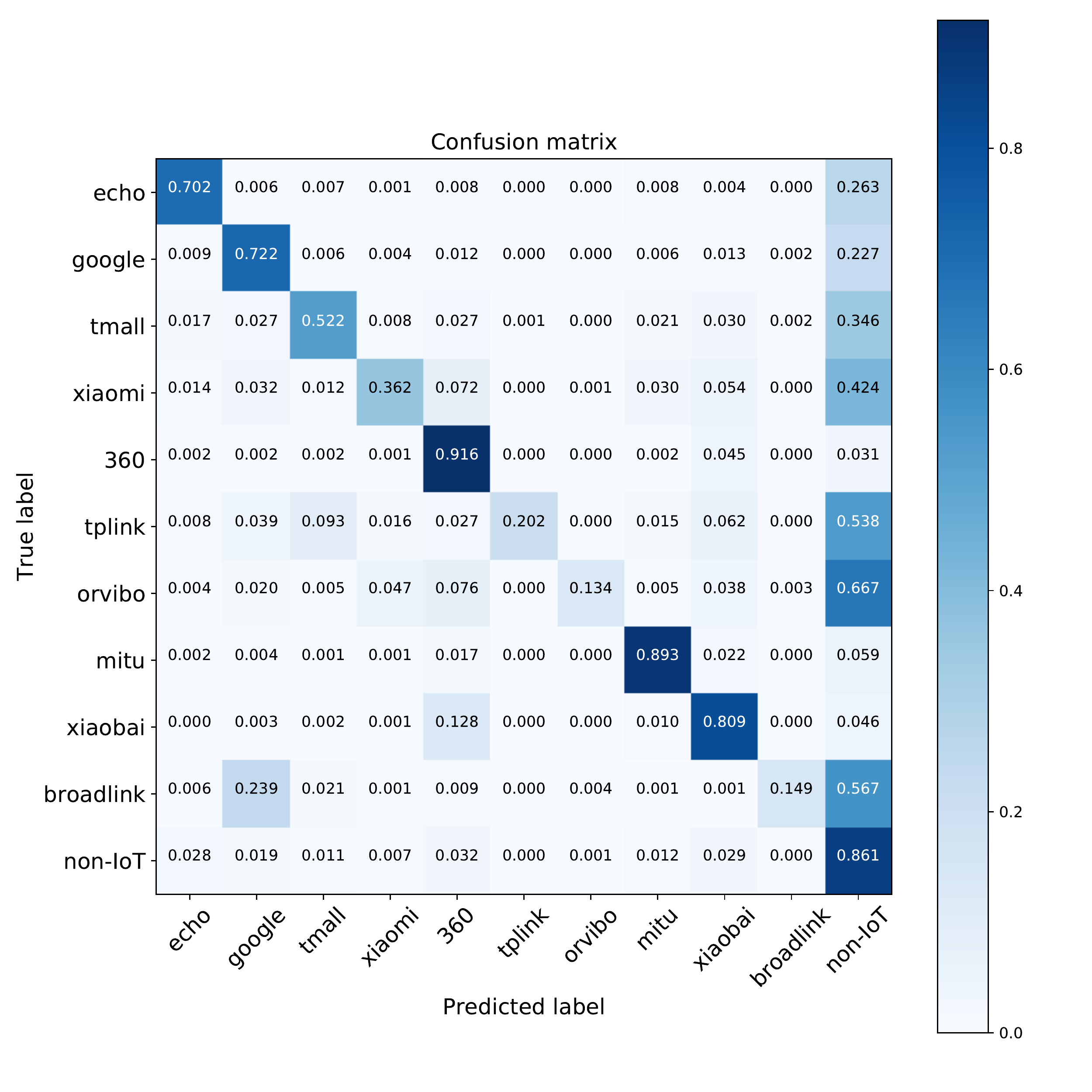} 
    	\caption{Confusion matrix of BLSTM (noisy+VPN).}
    	\label{fig:cm_blstm_noisy_vpn}
    	    	\vspace{-0.2in}

    \end{minipage}
\end{figure*}

Figure~\ref{fig:cm_rf_noisy_napt} and Figure~\ref{fig:cm_rf_noisy_vpn} show the performance of random forest in NAPT and VPN configurations in noisy environment. Figure~\ref{fig:cm_blstm_noisy_napt} and Figure~\ref{fig:cm_blstm_noisy_vpn} show the performance of bidirectional LSTM in NAPT and VPN configurations within noisy environment.


\end{appendices}
	
	%

\end{document}